\documentclass[fleqn,usenatbib]{mnras}

\usepackage{newtxtext,newtxmath}

\usepackage[T1]{fontenc}
\usepackage{datatool}
\usepackage{comment} 

\usepackage{graphicx}	
\usepackage{amsmath}	
\usepackage{amssymb}	
\newcommand\textlcsc[1]{\textsc{\MakeLowercase{#1}}}

\title[GES and the dual thick disc/halo]{Sausage \& Mash: The dual origin of the Galactic thick disc and halo from the gas-rich Gaia-Enceladus-Sausage merger}

\author[R. J. J. Grand et al.]{\parbox[t]{\textwidth}{
Robert J. J. Grand,$^{1}$\thanks{E-mail: grand@mpa-garching.mpg.de}
Daisuke Kawata$^2$, Vasily Belokurov$^3$, Alis J. Deason$^4$, Azadeh Fattahi$^4$, Francesca Fragkoudi$^1$, Facundo A. G\'{o}mez$^{5,6}$, Federico Marinacci$^7$, R\"udiger Pakmor$^1$}
\\\\
$^{1}$Max-Planck-Institut f\"{u}r Astrophysik, Karl-Schwarzschild-Str. 1, 85748 Garching, Germany\\
$^{2}$Mullard Space Science Laboratory, University College London, Holmbury St. Mary, Dorking, Surrey, RH5 6NT, UK\\
$^3$Institute of Astronomy, Madingley Rd, Cambridge, CB3 0HA\\
$^4$Institute for Computational Cosmology, Department of Physics, University of Durham, South Road, Durham DH1 3LE, UK\\
$^5$Instituto de Investigaci\'on Multidisciplinar en Ciencia y Tecnolog\'ia, Universidad de La Serena, Ra\'ul Bitr\'an 1305, La Serena, Chile\\
$^6$Departamento de Astronom\'ia, Universidad de La Serena, Av. Juan Cisternas 1200 Norte, La Serena, Chile\\
$^7$Department of Physics \& Astronomy, University of Bologna, via Gobetti 93/2, 40129 Bologna, Italy
}

\date{Accepted XXX. Received YYY; in original form ZZZ}

\pubyear{2019}

\hypersetup{draft}
\begin{document}
\label{firstpage}
\pagerange{\pageref{firstpage}--\pageref{lastpage}}
\maketitle

\begin{abstract}
We analyse a set of cosmological magneto-hydrodynamic simulations of the formation of Milky Way-mass galaxies identified to have a prominent radially anisotropic stellar halo component similar to the so-called ``Gaia Sausage'' found in the Gaia data. We examine the effects of the progenitor of the Sausage (the Gaia-Enceladus-Sausage, GES) on the formation of major galactic components analogous to the Galactic thick disc and inner stellar halo. We find that the GES merger is likely to have been gas-rich and contribute 10-50$\%$ of gas to a merger-induced centrally concentrated starburst that results in the rapid formation of a compact, rotationally supported thick disc that occupies the typical chemical thick disc region of chemical abundance space. We find evidence that gas-rich mergers heated the proto-disc of the Galaxy, scattering stars onto less-circular orbits such that their rotation velocity and metallicity positively correlate, thus contributing an additional component that connects the Galactic thick disc to the inner stellar halo. We demonstrate that the level of kinematic heating of the proto-galaxy correlates with the kinematic state of the population before the merger, the progenitor mass and orbital eccentricity of the merger. Furthermore, we show that the mass and time of the merger can be accurately inferred from local stars on counter-rotating orbits.
\end{abstract}




\section{Introduction}

Since the pioneering work of \citet{Eggen+Lynden-Bell+Sandage62}, the stellar halo of the Milky Way is considered to provide the crucial archaeological record from which the formation history of our Galaxy may be reconstructed. \citet{Eggen+Lynden-Bell+Sandage62} found an apparent correlation between the orbital eccentricity ($e$) and the metallicity of halo stars, with a concentration of high-$e$ and low-[Fe/H] stars and another concentration of more disc-like, low-$e$ and high-[Fe/H] stars. They interpreted this correlation as the signature of a rapid collapse of the Milky Way at an early epoch. Later, using a larger data-set combined with data from the European Space Agency's Hipparcos \citep{Perryman+97} mission and spectroscopic surveys, \citet{Chiba+Beers00} showed that there is no such strong correlation between $e$ and [Fe/H]. However, in their paper, there is a concentration of stars around $e\sim0.9$ and [Fe/H]$\sim-1.7$. \citet{Brook+03} analysed the phase space distribution of these high-$e$ and low-[Fe/H] stars, and concluded from a comparison of the observed phase space distribution to that of a numerical simulation that these stars originated from a dwarf galaxy accreted onto the Milky Way. Hence, the high-$e$ and low-[Fe/H] concentration of stars in \citet{Eggen+Lynden-Bell+Sandage62} is not evidence of a rapid, early collapse, but rather evidence of satellite accretion. Further hints of the significant contribution of satellite accretion in halo stars are observed in, for example, 2MASS data \citep[e.g.][]{Ibata2002,Majewski2003}, SDSS data \citep[e.g.][]{Yanny2000,Newberg2002,Belokurov2006, BZB08,Smith2009}, SEGUE spectroscopy \citep[e.g.][]{CBL08,SRA09,Xue2011} and the detailed abundance patterns of halo stars from high-resolution spectroscopy data (\citealt{Roederer2009}, \citealt{Nissen+Schuster10}, \citealt{ICA10}, \citealt{Roederer2014}). This idea has been recently affirmed with the {\it Gaia} data, which has revealed that the local stellar halo contains a significant amount of debris from an early merger of a satellite galaxy dubbed the "{\it Gaia} Sausage" \citep{BEE18} and "{\it Gaia}-Enceladus" \citep{HBK18}. We will refer to this significant early merger as the Gaia-Enceladus Sausage (GES). \citet{Mackereth+Schiavon+Pfeffer+19} analysed the orbital eccentricities of stars and confirmed that the GES stars are of high-$e$. GES stars are found also to have noticeably different chemical compositions compared to in-situ stars commonly interpreted as thick disc or halo stars, characterised in particular by low-[Fe/H] and high-[$\alpha$/Fe] \citep{HBK18,Haywood+DiMatteo+Lehnert+18}. 

Using the second data release (DR2) of the {\it Gaia} mission \citep{Gaia+Brown+Vallenari+18}, \citet{Haywood+DiMatteo+Lehnert+18} noted that the low- and high-[$\alpha$/Fe] sequence of the halo stars observed by \citet{Nissen+Schuster10} corresponds to the blue and red sequences of the main-sequence in the Hertzsprung-Russel Diagram (HRD) for kinematically identified halo stars. They suggested that one of these populations was formed in-situ in the Milky Way progenitor and that the other population formed in other galaxies accreted onto the Milky Way, which is dominated by GES stars. Combining the {\it Gaia}~DR2 data with  SDSS APOGEE data, \citet{DiMatteo+Haywood+Lehnert+18} found that the high-[$\alpha$/Fe] sequence of the halo stars have high-[Fe/H] similar to the thick disc population, and that they are kinematically connected to each other. \citet{Gallart+Bernard+Brook+19} analysed the age distribution of the blue and red sequences in the HRD, and found that the age of these two halo populations are similar, but the blue sequence is more metal poor than the red sequence. Comparing with a numerical simulation, they concluded that the blue sequence likely originated from an accreted dwarf galaxy, i.e. the GES, and the red sequence comprises the stars formed in-situ in the proto-disc of the Milky Way that were later dynamically ejected into the halo as a result of the GES merger. This scenario explains the higher metallicity of the red sequence compared to the blue sequence: chemical evolution progresses more rapidly in larger galaxies. \citet{BSF20} named these ``kicked out'' halo stars the "Splash", and analysed the extensive data-set that combined {\it Gaia}~DR2 with data from several different spectroscopic surveys, including APOGEE~DR14, LAMOST~DR2, RAVE~DR5, Gaia-ESO~DR3, GALAH~DR2 and SEGUE, which is compiled and value added in \citet{Sanders+Das18}. They thoroughly analysed the chemical abundances, kinematics and ages of the stars in the halo and thick and thin discs. They argued that there are three kinematically different components: the Splash; the thick disc and the thin disc, for stars with [Fe/H]$>-0.7$, and that these components occupy different regions in the rotation velocity, $V_{\phi}$, and [Fe/H] plane. They found also that the age of the Splash component is as old as the other more metal poor halo stars, which are thought to have formed in smaller accreted galaxies, including the GES. \citet{BSF20} compared their observational data with the state-of-the-art cosmological numerical simulations of the Milky Way-like galaxies, \textlcsc{Auriga} \citep{GGM17,GHF18} and \textlcsc{Latte} \citep{Sanderson+Wetzel+Loebman+18}, and found evidence that the Splash component is produced from a merger that dynamically heats stars from the Milky Way proto-disc onto halo-like orbits \citep[see also][]{ZWB09,FMC11,MFC12,MGG19,BHT19}.

These recent studies suggest that the Splash component is connected to the thick disc. However, a remaining question is how the thick disc is related to the event of the merger with the GES progenitor. \citet{HBK18} discussed that the impact of the GES merger is significant enough to heat the proto-disc of the Milky Way and create the thick disc, under the assumption that a thinner proto-disc existed before the merger event. \citet{BSF20} alluded that dwarf galaxies at early epochs must be gas-rich and therefore an ample amount of gas is fed into this merger event and creates yet another population of stars distinct from the Splash. Based on a Cold Dark matter (CDM)-based "semi-cosmological" N-body/Smoothed Particle Hydrodynamics (SPH) simulation, \citet{Brook+Kawata+Gibson+Freeman04} first showed that the mergers of smaller, gas-rich galaxies are naturally expected at an early epoch of a Milky Way-like disc galaxy in the CDM universe. There is evidence that such gas-rich mergers are capable of inducing the formation of a thick disc \citep[see also][]{Robertson+Bullock+Cox+06}, and subsequent build-up of the thin disc \citep{Brook+Kawata+Martel+06}, which has been confirmed by more sophisticated and higher-resolution cosmological simulations of Milky Way-like galaxies \citep[e.g.][]{BiKW12,BSG12,SBR13,MCM14b,GBG18}. Moreover, gas-rich mergers at early times appear to naturally explain the high-velocity dispersion and high-[$\alpha$/Fe] abundances of thick disc stars \citep{Brook+Kawata+Richard+07}. Thus, the stellar population formed during the merging of the GES progenitor system is likely a significant component of the Milky Way that must be understood in relation to the Splash in order to decipher the chemodynamical properties of our Galaxy.  

Using the state-of-the-art \textlcsc{Auriga} suite of cosmological, magnetohydrodynamical zoom-in simulations of Milky Way-like galaxies, we show that a GES-like merger at high redshift is likely to be gas-rich and induce a starburst. We study how this merger impacts the proto-disc stellar components formed before the merger and the subsequent generation of stars formed from the merger-induced starburst, which contributes to the formation of the metal-rich stellar halo and the thick disc. Although the simulations are not intended to reproduce all the properties of the Milky Way, they provide physical insight into how a GES-like merger affected the early stages of disc galaxy formation. This is particularly useful because neither the properties of the GES merger nor the proto-disc of the Milky Way before the merger are known. We investigate how the impact of the merger depends on the mass of the GES progenitor and the kinematic properties of the proto-disc prior to the GES merger for the range of analogous systems of the \textlcsc{Auriga} suite. We further predict that the age distribution of retrograde Splash stars provide an accurate estimation of the GES merger time under the assumption that the Milky Way had no significant mergers afterwards and provided reliable age-estimates for these stars can be measured.

In section~\ref{sims}, we describe the simulations on which we base our analysis. In section~\ref{results}, we present the results of our analysis, and discuss and conclude our findings in section~\ref{conclusions}.

\section{Simulations}
\label{sims}

We analyse a set of high resolution, cosmological magneto-hydrodynamical simulations for the formation of the Milky Way \citep[from the \textlcsc{Auriga} project,][]{GGM17,GHF18}. These simulations are taken from the ``level 4'' resolution suite ($\sim 5 \times 10^4$ $\rm M_{\odot}$ per baryonic element; physical softening length of $369$ pc after $z=1$). All haloes range between $1$-$2\times 10^{12}$ $\rm M_{\odot}$ in total mass ($M_{200}$), which we define as the mass contained inside the radius at which the mean enclosed mass volume density equals 200 times the critical density for closure. Each halo was initially selected from the $z=0$ snapshot of a parent dark matter only cosmological simulation of comoving periodic box size 100 Mpc, with the standard $\Lambda$CDM cosmology. The adopted cosmological parameters are $\Omega _m = 0.307$, $\Omega _b = 0.048$, $\Omega _{\Lambda} = 0.693$ and a Hubble constant of $H_0 = 100 h$ km s$^{-1}$ Mpc$^{-1}$, where $h = 0.6777$, taken from \citet{PC13}. At $z=127$, the resolution of the dark matter particles of this halo is increased and gas is added to create the initial conditions of the zoom simulation, which is evolved to present day with the magneto-hydrodynamics code \textlcsc{AREPO} \citep{Sp10}. 

The simulations include a comprehensive self-consistent\footnote{The nomenclature in the literature to describe merely that a model is consistent with itself.} galaxy formation model \citep[][]{GGM17}, which treats many relevant processes, including gravity, gas cooling, star formation, mass and metal return from stellar evolutionary processes, energetic stellar and AGN feedback, and magnetic fields. Below, we summarise certain aspects of the model that are particularly pertinent to Galactic archaeology, namely the nucleosynthesis of heavy elements and their redistribution and mixing in galactic gas.

Star particles are able to form from gas denser than 0.11 atoms $\rm cm^{-3}$ \citep{SH03}, and represent single stellar populations (SSPs) of a given age, metallicity and mass, distributed according to a \citet{C03} initial mass function. Mass and metal loss rates from AGB stars \citep{K10} and Type Ia supernovae \citep[yields from][]{TAB03,THR04} are calculated, the latter using a delay time distribution \citep{MMB12}, and are distributed into neighbouring gas cells. Type II supernovae are modelled with a phenomenological wind model \citep[see for example][]{VGS13}, whereby a ``wind particle'' carries mass, metals \citep[yields from][]{PCB98} and energy away from the production site \citep[see][for a complete description of the \textlcsc{Auriga} model]{GGM17}. Turbulence and mixing is accurately followed in \textlcsc{AREPO}'s quasi-Lagrangian moving mesh scheme on scales above the cell size, which can be as small as $\sim 10$ pc in the densest regions. This naturally follows the mixing of metals in and between different gas phases in contrast to simulations that employ SPH and Meshless-Finite-Mass (MFM) schemes \citep[e.g.][]{H17}, in which the transfer of metals between resolution elements requires some form of additional metal diffusion scheme with somewhat arbitrary diffusion coefficients. 

The \textlcsc{AURIGA} galaxy formation model has been shown to produce realistic spiral disc galaxies that are broadly consistent with a number of observations including star formation histories, stellar masses, sizes and rotation curves of Milky Way-mass galaxies \citep{GGM17}, the distribution of HI gas \citep{MGP16}, the stellar halo properties of local galaxies \citep{MGG19}, stellar disc warps \citep{GWG16}, the properties and abundance of galactic bars \citep{Fragkoudi+Grand+Pakmor+19} and bulges \citep{GMG19}, the number and SFHs of Milky Way dwarf satellite galaxies \citep{SGG17,DNF19}, and the properties of magnetic fields in nearby disc galaxies \citep{PGG17,PGP18}. Furthermore, they are able to reproduce striking features of the Milky Way's stellar distribution, in particular: the Monoceros Ring \citep{GWM15}; the chemical thin/thick disc dichotomy \citep{GBG18}; the boxy-peanut bulge \citep{Fragkoudi+Grand+Pakmor+19}; and the radially anisotropic \emph{Gaia} ``Sausage'' \citep{FBD19}. Importantly, the galaxy formation model has been validated in large cosmological box simulations \citep[e.g.][]{PNH18,NPS18}, reproducing many of the global properties of the observed galaxy population, such as the galaxy stellar mass function, galaxy sizes, the cosmic star formation rate density and galaxy morphological mix. The combination of a globally successful model with the ability to resolve internal galactic structures, such as bars and spiral arms, is an extremely important and desirable trait for cosmological-zoom simulations, and makes the \textlcsc{Auriga} suite ideal to study Galactic dynamics within the full cosmological setting.

In this paper, we analyse a total of 22 simulations taken from the Auriga simulations \citet{GGM17}: 9 of these are highlighted in \citet{FBD19} to have a similar fraction of highly radially anisotropic stellar halo stars to the recently discovered GES feature in the Milky Way's stellar halo \citep{BEE18,HBK18}. The additional 13 simulations have also a significant merger event at similar epochs to the ``Sausage''-like haloes, but produce less radially-anisotropic velocity distributions for halo stars. We include these haloes to examine the orbital eccentricity of the GES-like merger and its effect on thick disc and halo formation. Six of the simulated haloes include Monte-Carlo (MC) tracer particles in their output. MC tracers are passive particles that are seeded in gas cells at the beginning of the simulations. A given tracer moves into neighbouring gas cells according to a probability proportional to the outward mass flux across a cell face normalised by the mass of the cell in which the tracer is located \citep[for more details we refer the reader to][and references therein]{GVZ19}. Thus, MC tracer particles enable us to statistically track the flow of gas over time and estimate the fraction of gas brought in by the GES progenitor that contributed to the formation of the stellar thick disc and halo. All simulations are run with the on-the-fly structure finder \textlcsc{SUBFIND} \citep{SWT01} which identifies gravitationally bound subhaloes and central haloes within groups identified by a friends of friends (\textlcsc{FOF}) algorithm \citep{DEF85}. In post-processing, we run a modified version of the LHaloTree algorithm to construct merger trees \citep[see][for details]{SGG17}, which we use to link the merger history of all subhaloes and identify the evolution of the GES-like merger in each simulation.

\begin{figure*}
\includegraphics[scale=1.2,trim={0 0 0 0},clip]{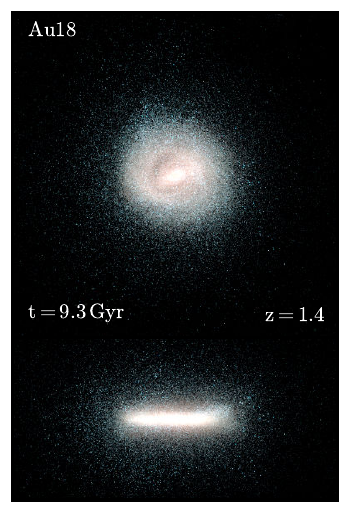}
\includegraphics[scale=1.2,trim={0 0 0 0},clip]{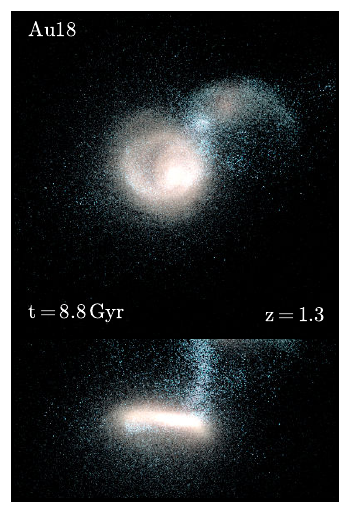}
\includegraphics[scale=1.2,trim={0 0 0 0},clip]{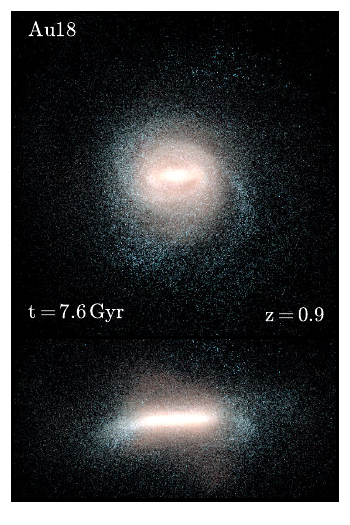}
\includegraphics[scale=1.2,trim={0 0 0 0},clip]{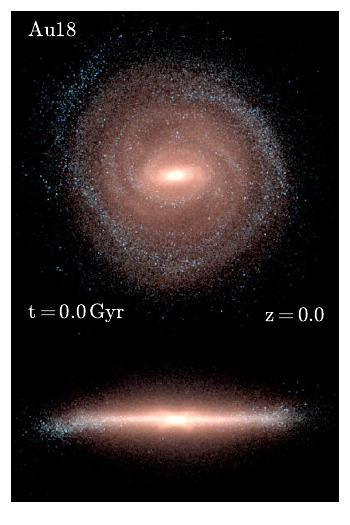}\\
\includegraphics[scale=1.2,trim={0 0.2cm 0 0},clip]{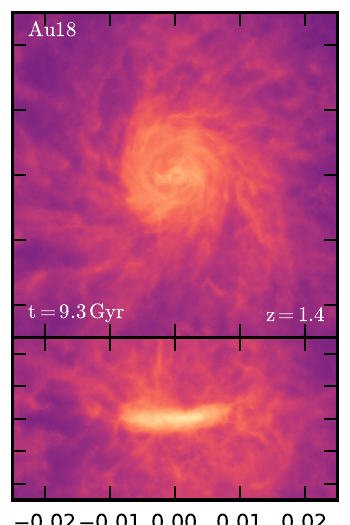}
\includegraphics[scale=1.2,trim={0 0.2cm 0 0},clip]{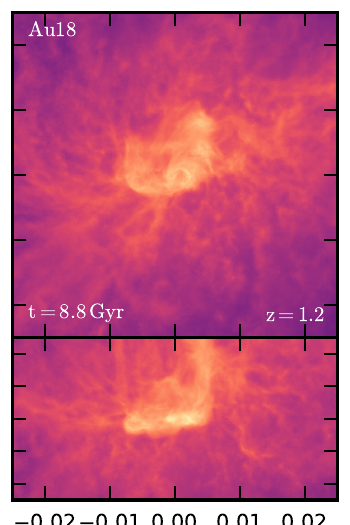}
\includegraphics[scale=1.2,trim={0 0.2cm 0 0},clip]{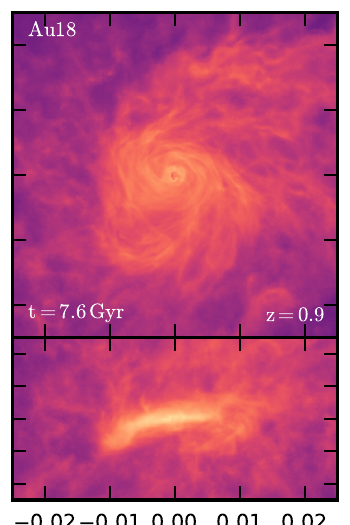}
\includegraphics[scale=1.2,trim={0 0.2cm 0 0},clip]{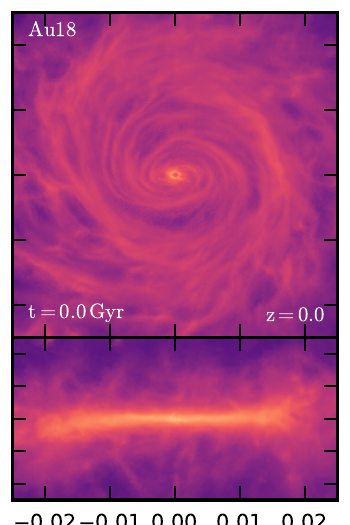}

\caption{Mock stellar light projections (upper panels) and gas density projections (lower panels) of the main galaxy in simulation Au 18. From left to right: $\sim 0.5$ Gyr before the GES merger; during the merger; $\sim 1.2$ Gyr after the GES merger; and redshift zero. }
\label{snaps}
\end{figure*}

\section{Results}
\label{results}

We first present detailed results from an individual simulated galaxy representative of our subset of Auriga galaxies. Our goal is to illustrate how the GES-like merger event shapes the stellar halo and thick and thin disc components observed at present day, with particular emphasis on their chemodynamical and morphological properties. Specifically, we will dissect the stellar populations formed before, during and after the GES-like merger event. In the second part of our analysis, we study the properties of these components across our subset of Auriga simulations in order to gain a statistical handle on the role of the GES-like merger on shaping the thick disc and halo. We conclude by making observational predictions for the mass and merger time of the GES merger. We stress that the results of this paper are based purely on numerical simulations, and to make the name concise, we simply say the GES component and GES merger to respectively describe the GES-like stellar component and the GES-like mergers identified in each simulation according to \citet{FBD19}. 

\begin{table*}
\caption{Masses of stellar populations and GES progenitor in the 9 simulations identified to have a radially anisotropic stellar halo in \citet{FBD19}. {\it Column 1:} the simulation name; {\it Columns 2-4:} the stellar mass of the proto-galaxy population (total, disc and spheroid); {\it Columns 5-7:} the stellar mass of the starburst population (total, disc and spheroid); {\it Columns 8-10:} the peak stellar mass (largest stellar mass attained by the GES progenitor in its evolutionary history prior to merging); the ratio of the GES peak stellar mass to the stellar mass of the Milky Way at the time the peak GES mass is attained (typically just before infall); and the peak gas mass of the GES progenitor. All masses are expressed as the logarithm of solar masses.}
\centering
\begin{tabular}{c c c c c c c c c c}
\\\hline
 & \multicolumn{3}{c|}{Proto-galaxy mass [$\rm log_{10}$]} & \multicolumn{3}{c|}{starburst mass [$\rm log_{10}$]} & \multicolumn{3}{c|}{GES progenitor} \\\hline

Halo & total & disc & spheroid &  total & disc & spheroid & $\rm log_{10}$ $M_{\rm *, GES}$ & $\rm M_{*,GES}/M_{*,MW}$ & $\rm log_{10}$ $M_{\rm gas,GES}$  
\\\hline
Au 5 & 10.3 & 10.1 & 10.0 & 9.8 & 9.7 & 9.2 & 9.6 & 0.11 & 10.3 \\
Au 9 & 9.9 & 9.6 & 9.6 &  9.8 & 9.5 & 9.5 & 9.3 & 0.12 & 10.4  \\
Au 10 & 10.3 & 10.1 & 9.8 & 9.7 & 9.6 & 9.2 & 9.0 & 0.03 & 10.0  \\
Au 15 & 10.0 & 9.7 & 9.7 & 9.5 & 9.4 & 9.0 & 9.4 & 0.15 & 10.3  \\
Au 17 & 9.9 & 9.7 & 9.5 & 9.8 & 9.6 & 9.3 & 9.1 & 0.03 & 9.7 \\
Au 18 & 10.3 & 10.2 & 9.8 & 9.9 & 9.7 & 9.2 & 9.2 & 0.04 & 10.1  \\
Au 21 & 9.9 & 9.6 & 9.5 & 9.8 & 9.5 & 9.4 & 9.8 & 0.4 & 10.5  \\
Au 22 & 10.3 & 10.0 & 9.9 & 9.8 & 9.7 & 9.1 & 8.2 & 0.02 & 9.6  \\
Au 26 & 9.9 & 9.4 & 9.7 & 10.1 & 9.7 & 9.8 & 10.0 & 0.65 & 10.7  
\\\hline
\end{tabular}
\label{t1}
\end{table*}

\subsection{A global, holistic view of the GES-thick disc-halo connection}

In this section, we present a detailed analysis of simulation Au 18, a representative case chosen to clearly illustrate the qualitative effects of the GES merger on the structure and chemo-dynamics of the thick disc and halo. In Fig.~\ref{snaps}, we show a series of snapshots around the event of the GES merger in simulation Au 18. Approximately 0.5 Gyr before the GES merger event takes place, the proto-galaxy has already formed a prominent stellar and gaseous disc, as illustrated in the first column of Fig.~\ref{snaps}. For convenience, we refer to the central galaxy before the GES-like merger as the proto-galaxy or the proto-disc. The stellar disc is visibly perturbed during the merger event (second column of Fig.~\ref{snaps}), and a gaseous bridge connects the GES progenitor to the main proto-galaxy, in which a trail of bright, blue stars form. This illustrates that the GES-like progenitor was a gas-rich system, which provided additional fuel for in-situ star formation. In fact, all the GES-like mergers in our simulations are gas-rich objects with a gas mass of order $\sim 10^{10}$ $\rm M_{\odot}$ (see Table~\ref{t1}). This is not surprising because MW-like galaxies are likely not sufficiently massive to develop the hot gas halo at early epochs \citep[e.g.][]{BGQ09,GBG18}, and the GES progenitor is massive enough to retain a significant fraction of its gas as it approaches and mergers with the central galaxy; ram-pressure stripping is not expected to strip all of the gas from sufficiently massive satellites. It follows that a gas-rich merger of significant mass on a quasi-radial orbit will produce significant star formation in what can be described as a merger-induced starburst. It is therefore necessary to consider the merger-induced star formation in order to fully understand the impact of the GES merger on the Milky Way. We will return to this important point later.

About 1 Gyr after the merger event (third column of Fig.~\ref{snaps}), the stellar light image indicates that some of the older, redder stars have scattered into a puffier configuration compared to the pre-merger time, contributing to the formation of the inner stellar halo together with accreted stars. However, the majority of stars remain in a component with a clear disc morphology. The younger, bluer stars, as well as the cold gas, are still settling into a quiescent disc - a phase which begins around this time and continues to grow inside-out and upside-down to form a radially extended thin disc, complete with a bar.

\begin{figure*}
\includegraphics[scale=1.5,trim={0 0 0.5cm 0.5cm},clip]{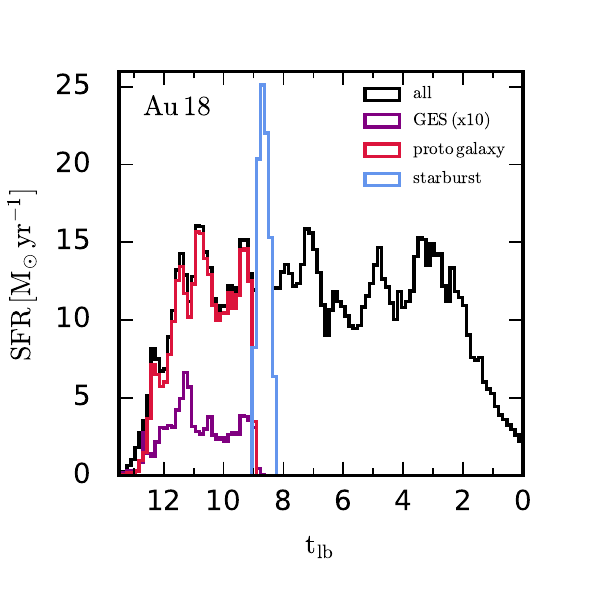}
\includegraphics[scale=1.5,trim={0cm 0 0.5cm 0.5cm},clip]{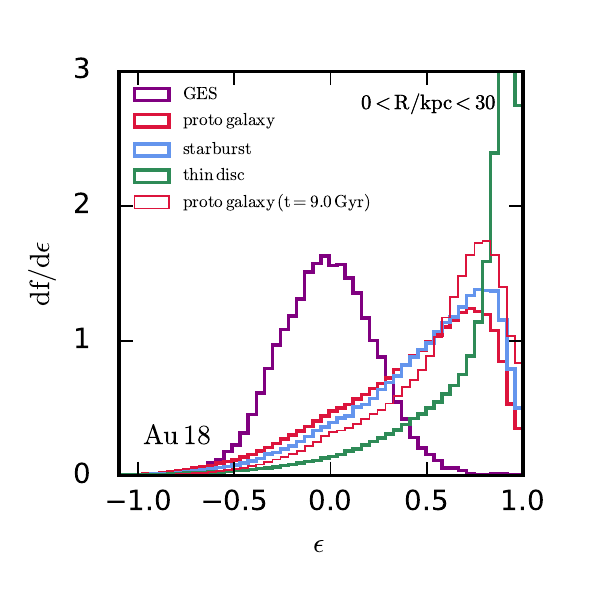}\\
\caption{The star formation history for Au 18. Here, we show the total SFH (black) and that of the GES progenitor (multiplied by 10, purple). The time of GES merger event can be inferred from the sharp drop in SFR of the latter, and is accompanied by a starburst that is driven by the gas-rich GES merger. We define the in-situ starburst population as the material formed during this burst in star formation (coloured blue). All stars that belong to the main halo before this starburst are identified as the pre-existing in-situ population which we refer to as the proto-galaxy population (coloured red). {\it Right panel}: The circularity distribution of the proto-galaxy population (red), starburst-population (blue), GES debris (purple) and thin disc (green) at redshift zero (thick histograms). In addition, we show the distribution of the proto-galaxy population before the GES merger (thin histogram), which has been kinematically heated by the merger. However, the proto-galaxy retains much of its rotation in a disc-like component. The starburst population is kinematically similar, but is on average (for our simulation sample) more rotationally dominated than the pre-existing population (see Secion.~\ref{simset}). All stars born after the merger (green) make up the thin disc, which is strongly rotating. The GES debris has very little net rotation.}
\label{sfh}
\end{figure*}

\begin{figure*}
\includegraphics[scale=0.5,trim={0 0 0 0},clip]{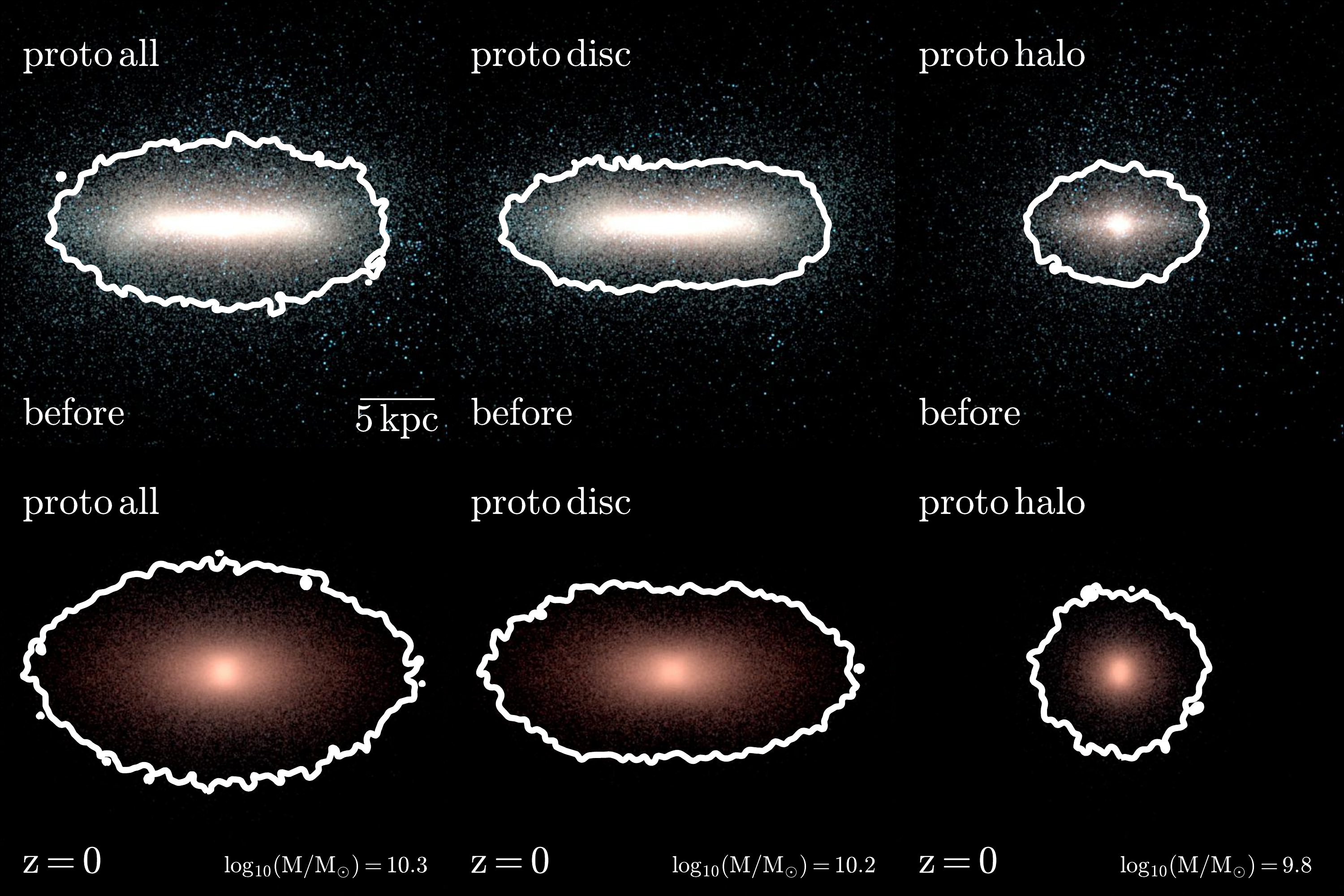}\\
\includegraphics[scale=0.5,trim={0 0 0 0},clip]{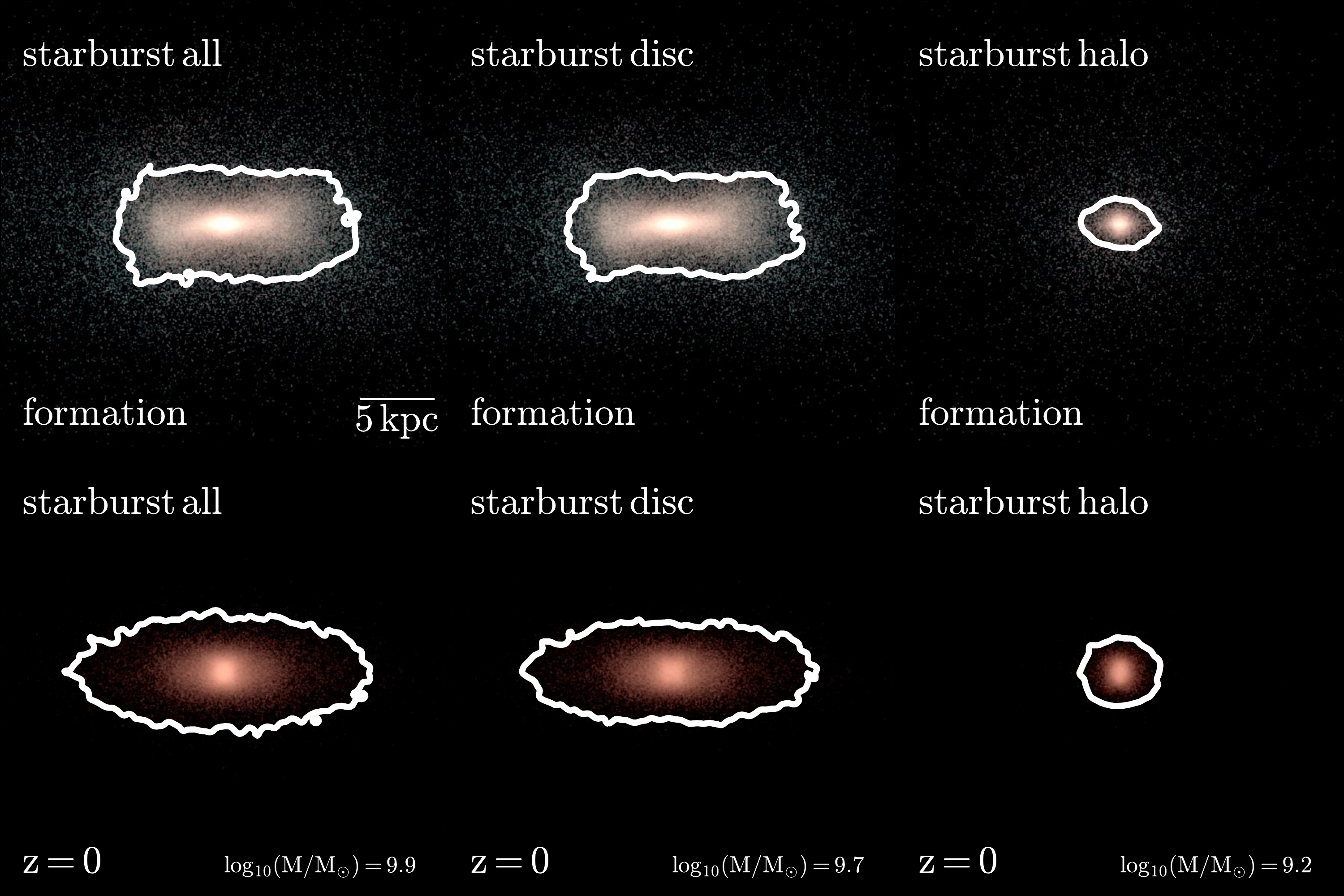}\\
\caption{Edge-on projections of the stellar light for the proto-galaxy (top 6 panels) and starburst (lower 6 panels) stellar populations. We kinematically decompose each population into a disc (middle column) and spheroid (halo/bulge; right-hand column) component. We show projections for the proto-galaxy stellar population both before the GES merger and at redshift zero. Before the GES merger, the proto-galaxy can be decomposed into a dominant stellar proto-disc, which the GES merger kinematically heats into a more puffy configuration as some proto-disc stars are scattered into the halo \citep[referred to as the Splash in][]{BSF20}. For the starburst component (lower 6 panels), we show projections shortly after the GES merger event and at redshift zero. The starburst stellar population is in general more compact than the proto-galaxy stars, which is mostly locked up in a rotating disc reminiscent of the Galaxy's thick disc. The remainder is found in a centrally concentrated spheroidal component. White contours are drawn to enclose the pixels with a projected surface mass density of at least $0.12$ $\rm M_{\odot}\,pc^{-2}$.}
\label{proj2}
\end{figure*}

\begin{figure*}
\includegraphics[scale=0.5,trim={0 0 0 0},clip]{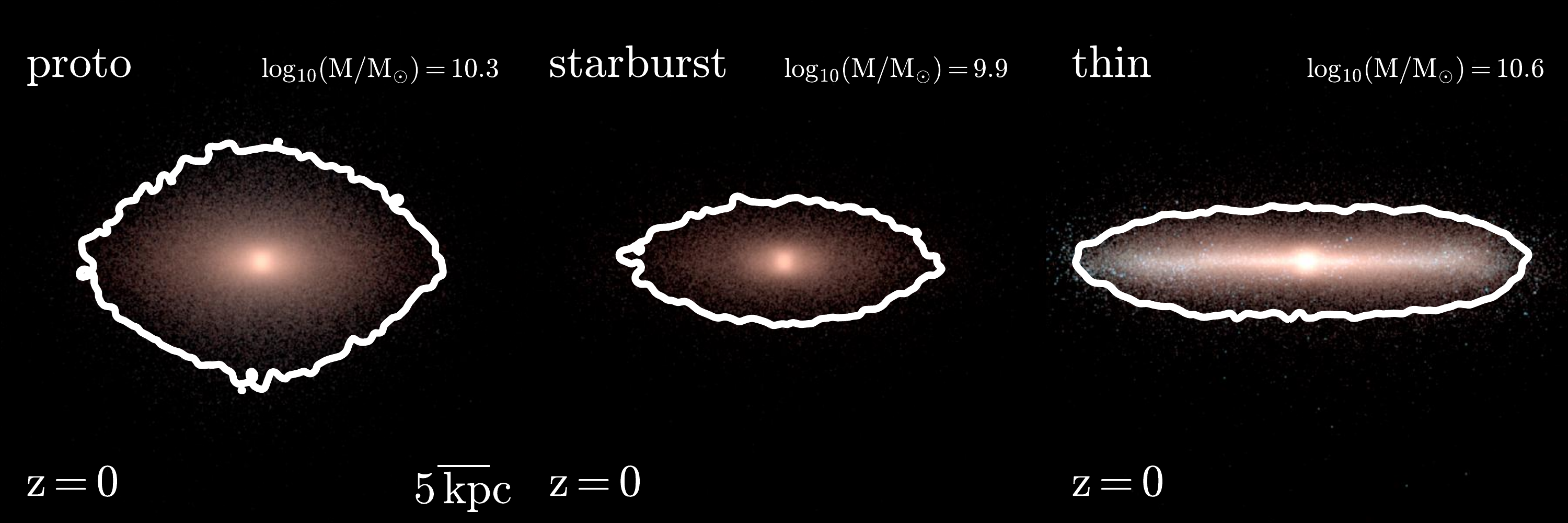}\\
\caption{Edge-on stellar light projections of the proto-galaxy, starburst and thin disc populations at redshift zero. The proto-galaxy is more extended than the starburst population. The thin disc population forms inside-out and upside-down from gas that settles into a star-forming disc after the GES merger event.}
\label{proj3}
\end{figure*}

\begin{figure*}
\includegraphics[scale=0.5,trim={0 -2.cm 0 0},clip]{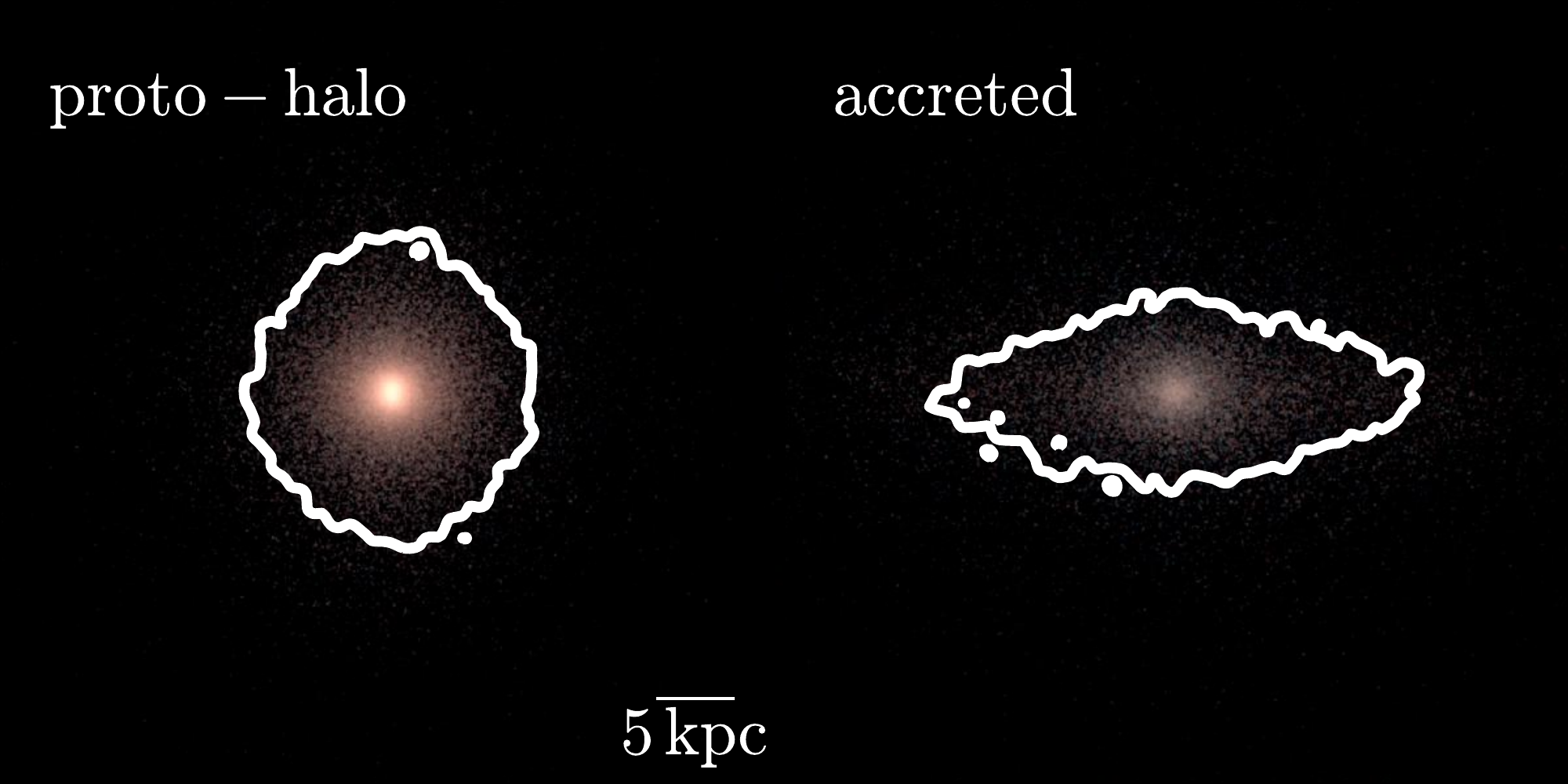}
\includegraphics[scale=1.2,trim={0.5cm 0.3cm 0 0},clip]{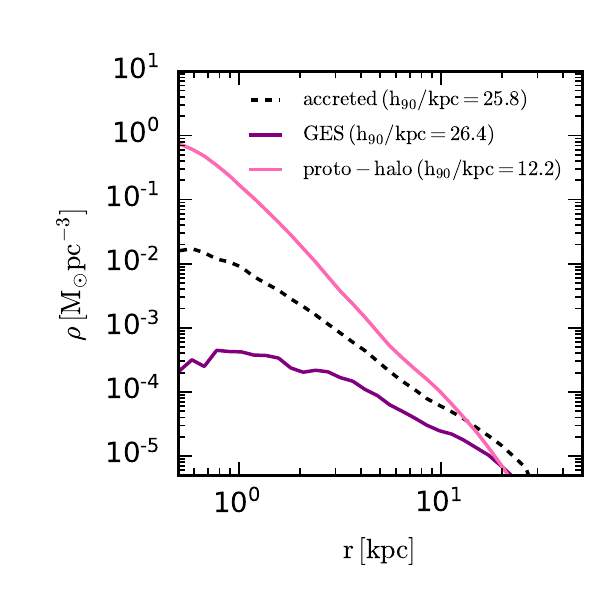}\\
\caption{Edge-on stellar light projections of the proto-halo (Splash) component (left panel) and the accreted component (middle panel). The spherical density radial profile for these components (right panel) show that the proto-halo is more compact compared to the accreted component; the former dominates over the latter within $\sim 20$ kpc of the Galactic centre. The radius enclosing $90\%$ of the mass of each component, $\rm h_{90}$, is indicated in the legend.}
\label{proj4}
\end{figure*}

In order to shed light on how this clearly defined merger event shaped the stellar populations seen at present day, we separate the stars formed before, during and after the merger. We illustrate this explicitly in the left panel of Fig.~\ref{sfh}, which shows the star formation history (SFH) of the simulated galaxy. To construct the SFH, we bin by age the star particles found inside a galactocentric radius of 30 kpc at redshift zero,  sum their initial masses\footnote{We sum their initial masses because mass loss from stellar evolutionary processes reduces the mass of stars over time.} in each age bin, and divide by the age bin-width.  

The SFH of the GES progenitor is also shown to be a factor of several below that of the MW progenitor. The truncation of the GES progenitor SFH at around 9 Gyr look back time marks the time of the merger, which is contemporaneous with a burst in star formation that peaks at roughly twice the SFR of the background SFR. This burst of star formation is caused by the increased compression of the gas in both merging gas-rich galaxies \citep{BSS18}, and represents a population of stars with a physically separate origin with respect to the stars that had formed before the merger event. Therefore, we define this population of stars formed during the merger as the {\it starburst} population, and designate in-situ stars that formed in the MW progenitor prior to the merger as {\it proto-galaxy} stars ( ex-situ/accreted stars are \emph{not} included in any component prefixed with ``proto'' in what follows). To be consistent with the nomenclature of \citet{BSF20}, we will use the term ``Splash'' to describe the proto-disc stars that are dynamically kicked (or splashed) out of the disc into the inner halo as a consequence of the GES merger. As explained below, not all proto-disc stars are splashed into the halo; some remain in a rotating (thick) disc-like configuration.

In the left panels of the first and second rows of Fig.~\ref{proj2}, we visualize the edge-on distribution of the stellar light of the proto-galaxy stellar population before the merger event takes place (at the time of the first column in Fig.~\ref{snaps}) and at redshift zero. The images show that the proto-galaxy is more spheroidal and radially extended at redshift zero than it was $\sim0.5$ Gyr before the GES merger event, which indicates that dynamical heating from the merger event has scattered star particles into the stellar halo (the Splash). In the left panels of the third and fourth rows in Fig.~\ref{proj2}, we show the edge-on distribution of stellar light of the starburst population $\sim 1$ Gyr after the merger event and at redshift zero. At both times, this component possesses a more compact and flattened morphology compared to the proto-galaxy stars. However this component is significantly thick, and extends to approximately solar radii.

Thus far we have discussed only the populations in a qualitative, general way, neglecting the possibility that each of these populations may be made from structurally (and kinematically) different components, such as a disc and halo/spheroid. We therefore quantify the disc and spheroidal components for each population by constructing a circularity distribution, shown in the right panel of Fig~\ref{sfh}. This figure shows the orbital circularity\footnote{Defined to be $\epsilon = \frac{L_z}{L_{z,\rm max}(E)}$, where $L_z$ is the $z$-component of angular momentum of a star particle and $L_{z,\rm max}(E)$ is the maximum angular momentum allowed for the orbital energy, $E$, of the star particle. For this definition, circular orbits in the direction of galactic rotation have $\epsilon =1$, counter-rotating orbits have $\epsilon <0$, and non-circular orbits have $\epsilon \sim 0$.} distributions of each population. The GES debris is peaked around a circularity of zero, reflecting its radial anisotropy with almost no net rotation. Both the starburst and proto-galaxy populations exhibit a significant degree of rotational support, though not as much as the thin disc stars that are strongly peaked around a circularity of one. The starburst and proto-galaxy distributions appear consistent with a rotating thick disc, with some evidence for a spheroidal halo-like component in the form of a mild bump around $\epsilon \sim 0$. 

In what follows, we decompose the starburst and proto-galaxy stellar populations into kinematically colder and hotter subsets, for which we adopt the nomenclature ``disc'' and ``spheroidal/halo'' components for brevity. We assign star particles to either of these components based on a probability defined by $p = y_{\rm halo} / y_{\rm tot}$, where $y=df/d\epsilon$ and $y_{\rm halo}$ is calculated to be the negative part of $df/d\epsilon$ mirrored around $\epsilon = 0$; everything with $\epsilon \leq 0$ is assigned to a spheroid (e.g. bulge, halo) component, whereas a star particle with $\epsilon >0$ is assigned to the disc if a uniformly drawn random number (between 0 and 1) is larger than the value $p(\epsilon)$, or to the spheroid/halo component otherwise. We note that material considered part of the stellar halo depends on the choice of cut on $\epsilon$. However, we emphasise that the purpose of applying such a ``definition'' is not meant to rigidly define distinct disc and halo components, but rather to gain intuition from the study of kinematically hotter and colder subsets of stars in order to facilitate our understanding of the proto-galaxy and starburst stellar populations as a whole. In fact, we will show in section ~\ref{cdr} that the chemo-dynamical properties of these components are rather smoothly connected.

We show the edge-on stellar light projections of the disc and halo components of both the starburst and proto-galaxy populations in the middle and right columns of Fig.~\ref{proj2}. As expected, it is mainly the kinematically-defined disc component of the proto-galaxy (proto-disc) that is dynamically heated. This is quantified in the right-hand panel of Fig.~\ref{sfh}, which shows that a fraction of rotating disc material is shifted into a more spheroidal, halo-like component (the Splash). The subdominant kinematically-defined proto-halo is less affected because it is already radially extended and kinematically hot. In contrast, the (subdominant) halo component of the starburst stellar population is very centrally concentrated because the violent gas-rich merger produces also a relatively small fraction of newborn stars on counter-rotating and/or radial orbits in the central region, owing to the compressible nature of gas. Thus, although the starburst driven by the gas-rich GES merger produces mainly thick disc-like, rotationally supported stars (see below and bottom-middle panels of Fig.~\ref{proj2}), the starburst adds also stars on radial/low-circularity counter-rotating orbits that contribute to a bulge/tail of the disc component. The properties of these stars can provide strong constraints on the mass of the GES progenitor, as discussed in \citet{Fragkoudi+Grand+Pakmor+19}. This may be also one of the mechanisms to form metal rich globular clusters, which are more centrally confined in the MW \citep[e.g.][]{Minniti95,Perez-Villegas+Barbuy+Kerber+19}. We note that even under the assumption that all star particles classified as part of the starburst ``halo'' component make up the central bulge (together with a fraction of the proto-halo), the bulge stellar mass in this simulation would be consistent with that of the Milky Way \citep[$<5\times10^9$ $\rm M_{\sun}$,][]{Shen+Rich+Kormendy+10,BHG16}.

The rapid formation and clear rotation present in the starburst population means that the gas in the proto-galaxy must be compressed, and the energy supplied by the merger event is rapidly dissipated. In addition, the violent GES merger does not allow the gas to settle into a kinematically thin disc, but rather sustains the star forming gas in a highly turbulent and kinematically hot state. Therefore, this is a natural process of forming the thick disc \citep{Brook+Kawata+Gibson+Freeman04}. Because our GES merger analogues are all gas-rich, it is natural to expect that some fraction of the starburst population is made from gas originating outside the proto-galaxy. We calculate this mass fraction by tracing the history of gas that formed these stars using ``tracer particles'' \citep{GVN13}. We stipulate that a gas element originated from the GES progenitor if it has at any time been bound to the ISM of the GES progenitor prior to the merger event \citep{GVZ19}. Of the 6 simulations for which tracer particles are available, we find that the GES progenitor provided between $10\%-50\%$ of the material that formed the starburst population. This suggests that a gas-rich GES progenitor is important for the rapid synthesis of the thick disc. 

\begin{figure*}
\includegraphics[scale=1.7,trim={1.cm 0 0.5cm 0},clip]{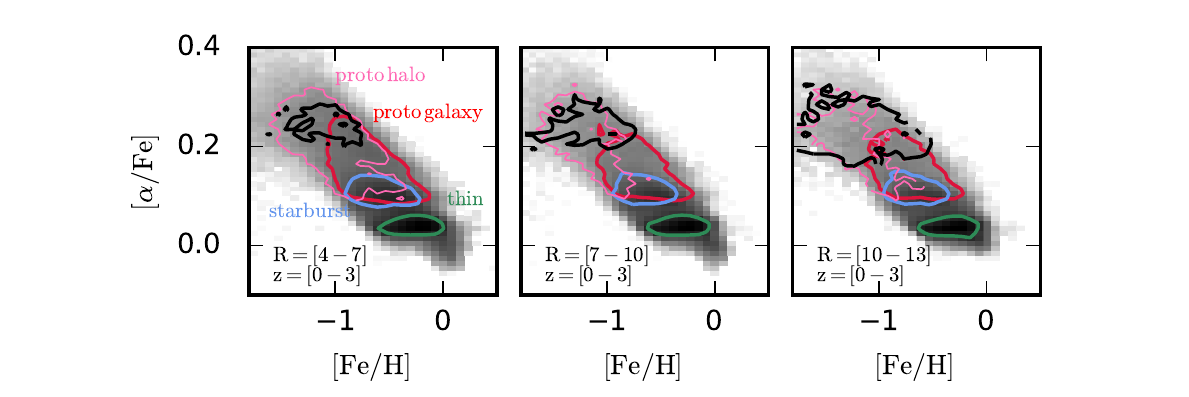}
\caption{The $\rm [\alpha/Fe]$ vs. $\rm[Fe/H]$ plane for logarithmic star counts at a range of cylindrical radii for simulation Au 18. Over-plotted in contours are the $25\%$ of the maximum star count per pixel for the proto-galaxy (including the Splash) population (red), in-situ starburst thick disc (blue), the in-situ thin disc (green) component and accreted stars (black). The starburst population lies at the metal-rich, alpha-poor end of the proto-galaxy, around the region expected for the traditional alpha-enhanced thick disc. The proto-galaxy extends from this region to higher alpha-abundances and lower metallicities where the proto-halo component (thin, pink contour) found. The metal-rich part of the proto-halo (Splash) does not extend as far out as the metal-poor halo, as a result of earlier kinematic heating of the older, metal-poor proto-galaxy. Accreted stars are found at low metallicities, and become more important relative to the proto-halo at larger radii, reflecting the density profiles shown in Fig.~\ref{proj4}. The thin disc is the most metal-rich, alpha-poor population, and forms a chemical bimodality with the alpha-enhanced thick disc.}
\label{aferz}
\end{figure*}

\begin{figure*}
\includegraphics[scale=1.7,trim={1.cm 0 0.5cm 0},clip]{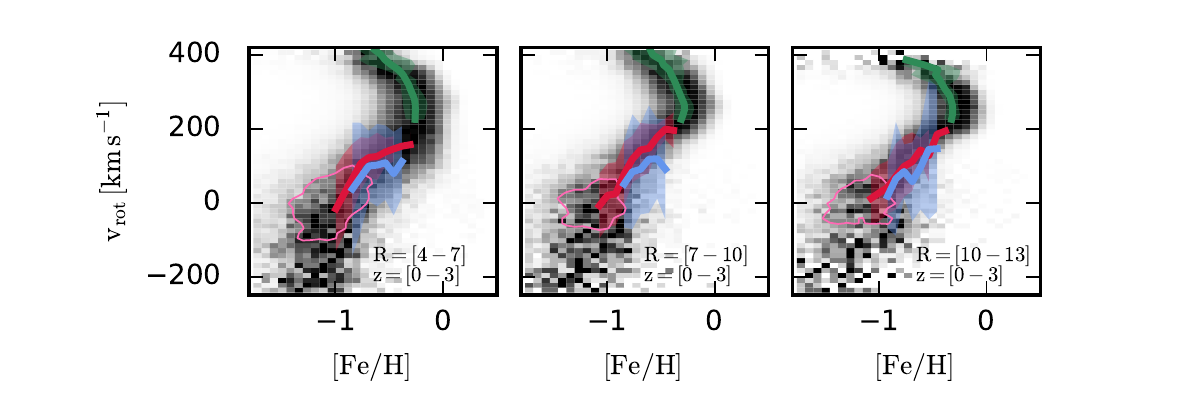}
\caption{As Fig.~\ref{aferz}, but in the rotational velocity-metallicity plane. The 2D histogram (grey scale) is row normalised. Over-plotted are mean rotation velocity-metallicity relations (solid curves) and scatter (shaded regions) for the proto-galaxy (red), starburst population (blue) and the thin disc (green). The pink contour delineates the locus of the proto-halo (Splash) population.}
\label{metvrot}
\end{figure*}

As described above, the middle panels of the upper two rows of Fig.~\ref{proj2} and the right panel of Fig.~\ref{sfh} show that the proto-disc is heated by the GES merger, but the majority of their stars maintain their prograde rotation, i.e. they are in a thick disc-like configuration. This indicates that some of the thick disc originates from the dynamically heated component of an originally colder proto-disc, which is similar to the classical thick disc formation scenario of a heated thin disc \citep{QHF93,DLQ11}. Note that the right panel of Fig.~\ref{sfh} shows that the proto-disc before the merger was not as cold as the thin disc at $z=0$. This is expected, because multiple minor mergers must have provided sufficient energy to prevent the formation of a kinematically cold disc before the GES merger. Therefore, this is a different scenario from the thin disc heating scenario for the formation of the thick disc. Nevertheless, the simulation indicates that the GES-like merger shapes the thick disc through two mechanisms: a gas-rich merger-induced turbulent starburst and the kinematical heating of the proto-disc, both of which are caused by the GES merger.

After the GES merger and formation of the starburst population, the thin disc begins to grow in earnest from a gas disc several kiloparsecs in radius, as seen in the lower-third panel of Fig.~\ref{snaps}. The thin disc grows inside-out and upside-down, such that by redshift zero \citep{Brook+Kawata+Gibson+Freeman04,Brook+Kawata+Martel+06,BiKW12,BSG12,SBR13,MCM14b,GBG18}, it is grown to approximately the same extent as the Splash population, as seen in Fig.~\ref{proj3}. 

Having illustrated the various in-situ components above, we place them in the context of the accreted population of stars to acquire an overall picture. Fig.~\ref{proj4} shows the edge-on projection of the proto-halo (in-situ by definition) with the accreted stellar population and their respective spherically averaged radial density profiles. In terms of stellar light, the proto-halo appears marginally rounder than the accreted stellar population. We remark that satellite accretion may produce ex-situ discs as well as halo stars \citep{RLA08,GGM17b}. The rightmost panel of Fig.~\ref{proj4} reveals that the proto-halo is far more compact than the accreted stellar distribution and dominates the stellar halo mass inside radii of $\sim 20$ kpc\footnote{We note that these statements are generally qualitatively true for our simulations, although the profile shapes of each component and the radius at which they overlap varies between simulations \citep[see][for a detailed investigation of the \textlcsc{AURIGA} stellar haloes]{MGG19}. Although these results provide a good qualitative picture, we emphasise that none of our simulated galaxies were tailored to match the Milky Way's stellar halo, therefore the simulated stellar halo properties (for example the fraction of accreted stellar halo) are likely not quantitatively identical to the Milky Way.}. Interestingly, the GES debris is even less compact than the total accreted star distribution, and traces a similar flattened morphology of all accreted stars as shown in the middle panel of Fig.~\ref{proj4}. This has important implications for future searches for GES debris stars; we will present a detailed study of the GES debris properties in a forthcoming publication.

\subsubsection{The $\rm [Fe/H]$ - $\rm [\alpha/Fe]$ plane}
\label{cdr}
In this section, we analyse the chemo-dynamics of each component we identified in the previous section. In Fig.~\ref{aferz}, we plot the $\alpha$-element (which we take to be magnesium) abundance, $\rm [\alpha/Fe]$, as a function of metallicity, $\rm[Fe/H]$, for the star particles in the galaxy as a function of radius, at redshift zero. We superpose contours that enclose pixels that contain at least 25 per cent of the maximum number of stars per pixel for each population, and further show such contours for the proto-halo, which extends from $\rm[\alpha/Fe] \sim 0.3$, $\rm[Fe/H]\sim -2$ to $\rm[\alpha/Fe] \sim 0.1$, $\rm[Fe/H]\sim -0.5$ in the central regions, where it connects to and overlaps the rest of the proto-galaxy population. 

Fig.~\ref{aferz} shows that the metal-rich part of the proto-halo does not extend as far as the metal-poor part, which is consistent with measurements of Milky Way stars tentatively identified as part of the Splash population presented by \citet{BSF20}. This suggests that the metal-poor part of the proto-halo is composed of proto-disc stars that were kinematically heated by (gas-rich) mergers prior to the GES merger. Indeed, this is supported by the top-middle panel of Fig.~\ref{proj2}, which shows that extended halo-like stars exist already before the GES merger. This scenario is natural in the hierarchical model of galaxy formation in the $\Lambda$CDM paradigm, and leads us to predict that the in-situ halo component of the MW is not exclusively produced by the GES merger. However, the GES merger produces most of the metal-rich proto-halo population, i.e. the Splash. We remind the reader that, because we mainly focus on how the GES merger affected the distribution of stars born in the early Milky Way, these stellar populations comprise in-situ stars only. In Fig.~\ref{aferz}, we show also the distribution of the accreted component (see also Fig.~\ref{proj4}), including the GES debris, which illustrates that the proto-halo is similarly metal-poor compared to the accreted stars.

\begin{figure*}
\includegraphics[scale=1.2,trim={0.5cm 0 0.6cm 0.2cm},clip]{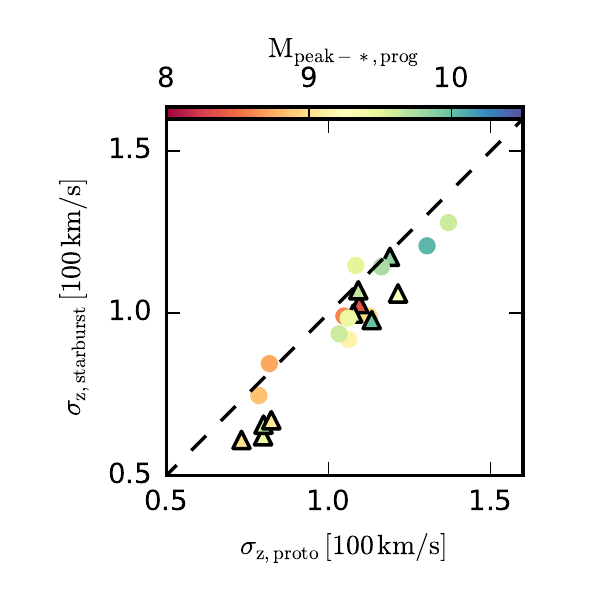}
\includegraphics[scale=1.2,trim={0.75cm 0 0.6cm 0.2cm},clip]{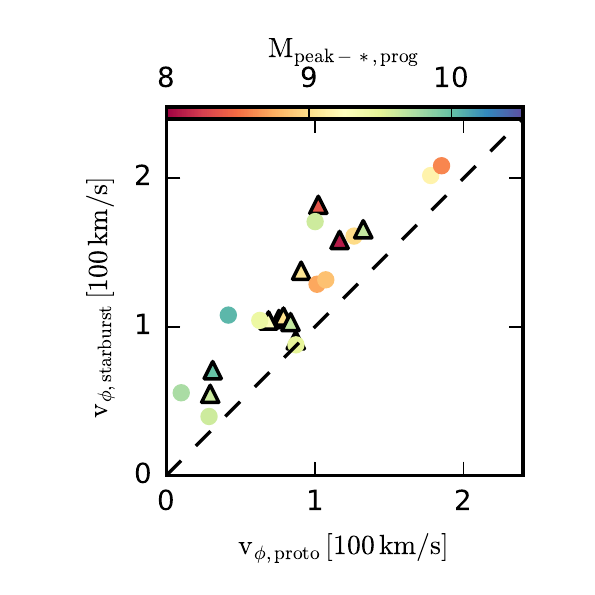} 
\includegraphics[scale=1.2,trim={0.75cm 0 0.6cm 0.2cm},clip]{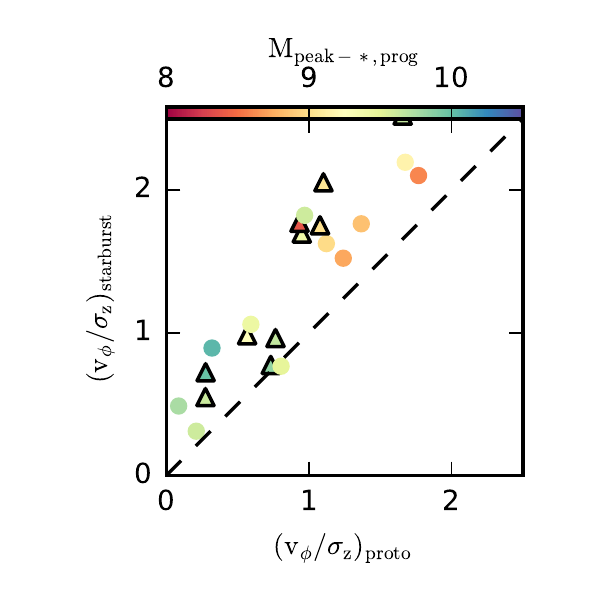}
\caption{Comparison of the kinematics (vertical velocity dispersion, $\sigma _z$, and rotational velocity, $\rm v_{\phi}$) of the proto-galaxy and the GES merger-induced starburst populations at redshift zero. The former population ranges from being kinematically similar to that of the latter, to kinematically hotter. In the majority of simulations, the starburst population carries more rotational support than the Splash population, either because the proto-disc was kinematically heated by the GES merger or because the proto-galaxy had not yet formed a disc at the time of merging (see Fig~\ref{heatsum}). Circles (triangles) indicate the simulations for which the GES progenitor orbit is more (less) eccentric than the median eccentricity for the simulation suite. }
\label{ksum}
\end{figure*}

The greater part of the proto-galaxy is in a rotating disc configuration that smoothly connects the halo to the old thick disc, and overlaps heavily with the metal-rich thick disc locus in chemical abundance space. This region is occupied also by the starburst population, which shows a compact disc morphology and thick disc-like rotation. Like the proto-disc, the locus of the starburst population in chemical abundance space is roughly constant with physical location, with a systematically high-[$\alpha$/Fe] compared to the chemically-defined low-[$\alpha$/Fe] thin disc. These properties are qualitatively consistent with those of the chemically-defined thick disc of the MW. The relatively metal-poor gas of the GES must be rapidly enriched and subsequently mixed during the starburst phase to produce the metal rich part of the thick disc population. In addition, this gas must rapidly dissipate the energy supplied by the GES merger in order to quickly settle into a metal-enriched, well-mixed disc configuration with rotational support, as seen in for example \citet{Robertson+Bullock+Cox+06} and \citet{Brook+Kawata+Martel+06}. Thus, the metal-rich part of the thick disc is composed of both the starburst population and proto-disc stars heated by the GES merger. The starburst and proto-disc components contribute to the metal-rich thick disc in roughly equal parts for this simulation\footnote{The starburst-disc to proto-disc mass ratio is about 1:2. However, the proto-disc extends to lower metallicities compared to the starburst population, therefore the mass in the metal-rich proto-disc material is comparable to that of the starburst in this region of abundance space.}, which is typical for the ``Sausage-like'' simulations listed in Table~\ref{t1}. 

Interestingly, Fig.~\ref{aferz} shows evidence of a bimodal distribution similar to that of the Milky Way. The alpha-poor part of the dichotomy is made of thin disc material that forms after the gas dissipates energy and becomes centrally concentrated as a result of the GES merger. After the merger, relatively metal-poor gas (compared to the enriched centrally concentrated gas) accretes from the circum-galactic medium to form the thin disc and the chemical dichotomy in a manner similar to the ``shrinking disc'' formation pathway discovered by \citet[][]{GBG18} \citep[see also][]{CMG97,SSM19}.

\subsubsection{The rotation velocity-metallicity plane}

In Fig.~\ref{metvrot}, we show the rotation velocity as a function of metallicity divided spatially as in Fig.~\ref{aferz}. We see that the proto-halo component has little rotation at all locations within the galaxy, as per definition. However, there is an overall positive correlation between rotation velocity and metallicity in the innermost radial bin such that the metal-rich end of Splash-halo exhibits a mild degree of prograde motion. In our simulation, we have verified that this trend is explained as follows: the metal-poor proto-halo is formed prior to the GES merger owing to a series of minor mergers; the proto-disc proceeds to grow more massive and metal-rich until the GES merger dynamically scatters (mainly) intermediate-metallicity stars into the (Splash-) halo component. These stars retain some of their rotation and dominate the proto-halo in the inner regions of the galaxy, in contrast to the outer regions which are dominated by the metal-poor proto-halo.

Interestingly, the starburst-halo component, though subdominant in mass (see Table. 1 and Fig.~\ref{proj2}), extends the stellar halo population to metallicities of the thick disc component, which indicates that the GES merger pushes (hydrodynamically) star-forming gas out of the disc. From Fig.~\ref{snaps}, we also infer that the stars formed in the bridge of cold gas immediately preceding the merger may contribute to this in-situ, metal-rich halo, highlighting the complexity involved in the genesis of these components.

As inferred from earlier figures, the proto-disc and starburst-disc components occupy a common region in rotation velocity-metallicity space at all locations in real space. Their distribution follows a positive slope, which is often discussed to be associated with the inside-out formation of the thick disc in the context of an isolated galaxy formation scenario \citep{KAB17,SM17,Minchev+Matijevic+Hogg+19}. In our simulation, however, the positive relation between metallicity and rotation velocity is built up by mergers: as the proto-disc evolves and self-enriches prior to the GES merger, multiple smaller mergers impart dynamical heating such that the older, relatively metal-poor proto-disc stars are heated more relative to the younger, more metal-rich stars, thus establishing the positive trend in rotation velocity-metallicity space for this component. At the same time, the starburst-disc forms quickly in a more centrally bound region, and therefore is more chemically enriched and carries more rotation compared to the proto-disc. Thus, the dual origin of the thick disc from the proto-disc and the starburst material, i.e. the ``mash'', can explain the trend between rotation velocity and metallicity and confirms the schematic picture presented by \citet{BSF20}.

\begin{figure*}
\includegraphics[scale=1.25,trim={0.75cm 0.1cm 0.4cm 0.2cm},clip]{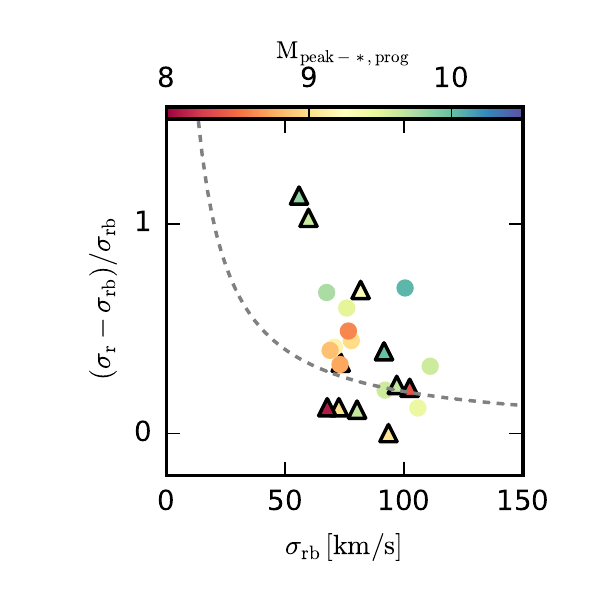}
\includegraphics[scale=1.25,trim={1.6cm 0.1cm 0.4cm 0.2cm},clip]{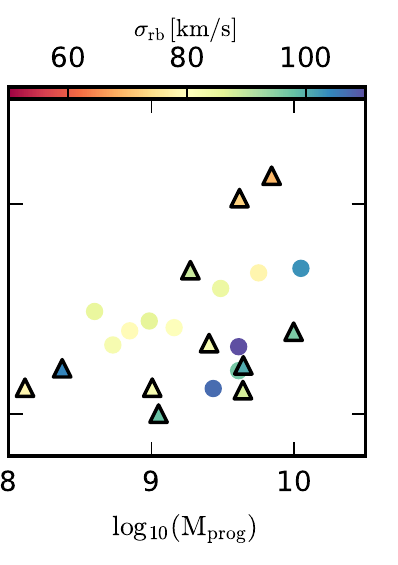}
\includegraphics[scale=1.25,trim={1.6cm 0.1cm 0.4cm 0.2cm},clip]{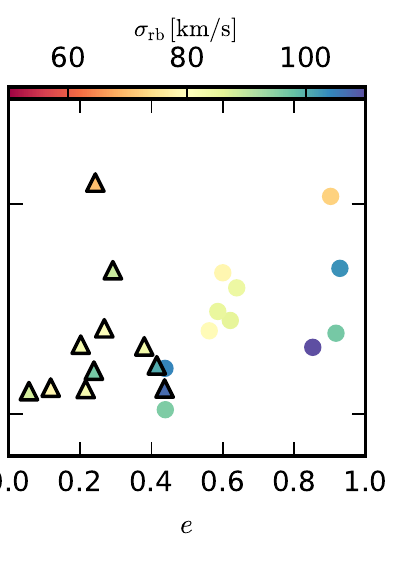}\\
\includegraphics[scale=1.25,trim={0.75cm 0 0.4cm 0.2cm},clip]{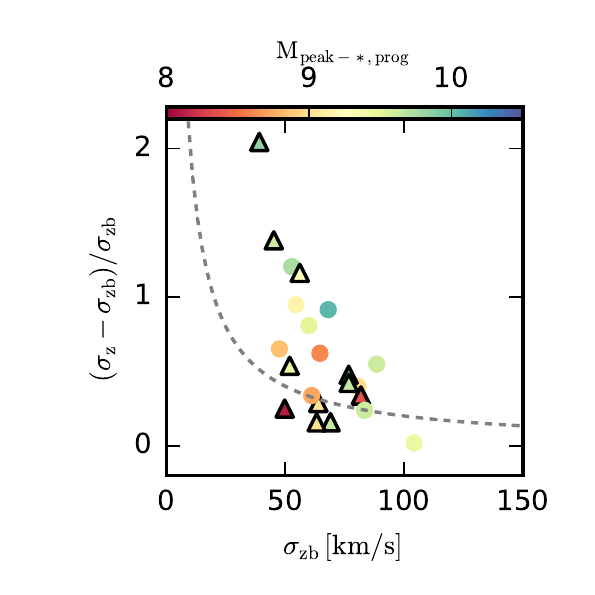}
\includegraphics[scale=1.25,trim={1.6cm 0 0.4cm 0.2cm},clip]{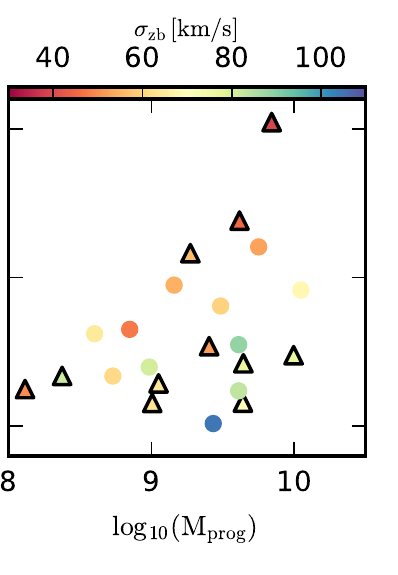}
\includegraphics[scale=1.25,trim={1.6cm 0 0.4cm 0.2cm},clip]{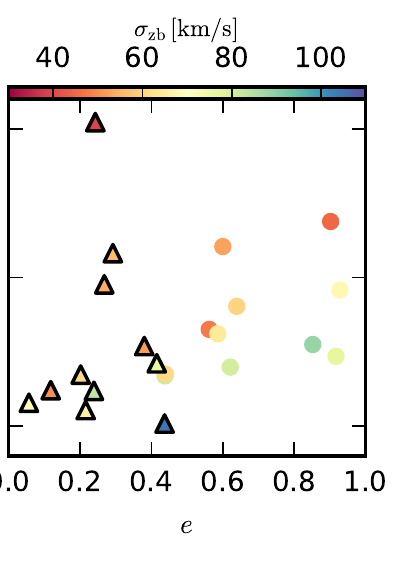}
\caption{Summary plot of the dynamics of the local, proto-galaxy stellar population just before the GES merger and at redshift zero. {\it Top panels}: the normalized change of the radial velocity dispersion as a function of: the radial velocity dispersion just before the impact, $\sigma _{\rm rb}$ (left); the maximum stellar mass attained by the GES progenitor before merging (middle); and the orbital eccentricity of the GES progenitor prior to merging (right). {\it Bottom panels}: as the top panels, but for the vertical velocity dispersion. Circles (triangles) indicate the simulations for which the GES progenitor orbit is more (less) eccentric than the median eccentricity for the simulation suite. The dashed curve in the left panels marks the locus for which the GES merger increases the velocity dispersion of the Splash population by 20 $\rm km\,s^{-1}$. A range of kinematic heating is seen in the sample: in some simulations the velocity dispersion is doubled, whereas in others they increase by less than 20 $\rm km\,s^{-1}$. The degree of fractional kinematic heating correlates with the initial kinematics (colder proto-discs preferentially experience more fractional kinematic heating relative to hotter proto-galaxy configurations), progenitor mass (more massive progenitors tend to provide more kinematic heating) and orbital eccentricity of the GES progenitor (more eccentric orbits tend to lead to more kinematic heating). We note that the \emph{absolute} kinematic heating is primarily a function of GES progenitor mass and orbital eccentricity (see Fig.~\ref{heatsumabs}).}
\label{heatsum}
\end{figure*}

The young thin disc occupies the high-rotation velocity and metal-rich part of the space, and shows a negative slope in the metallicity-rotation velocity relation throughout the disc. This is naturally explained by a negative radial metallicity gradient and epicycle motions of stars \citep[e.g.][]{AKC16}. Metal-poor stars have larger guiding centre radii (angular momentum) and they come to the radial region in Fig.~\ref{metvrot} only when they are at their peri-centre orbital phase. The rotation velocity peaks at a metallicity of $\rm[Fe/H]\lesssim -0.7$, which indicates the lowest metallicity in the thin disc stars.

\subsection{Simulation-wide trends}
\label{simset}

\begin{figure*}
\includegraphics[scale=1.25,trim={0.5cm 0 0.6cm 0.2cm},clip]{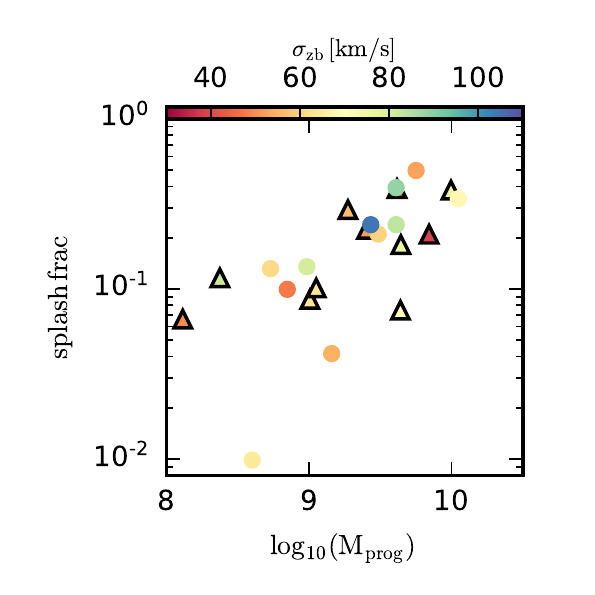}
\includegraphics[scale=1.25,trim={1.6cm 0 0.5cm 0.2cm},clip]{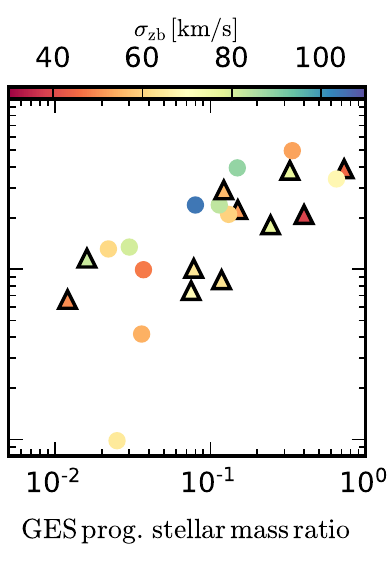}
\includegraphics[scale=1.25,trim={0.4cm 0 0.5cm 0.2cm},clip]{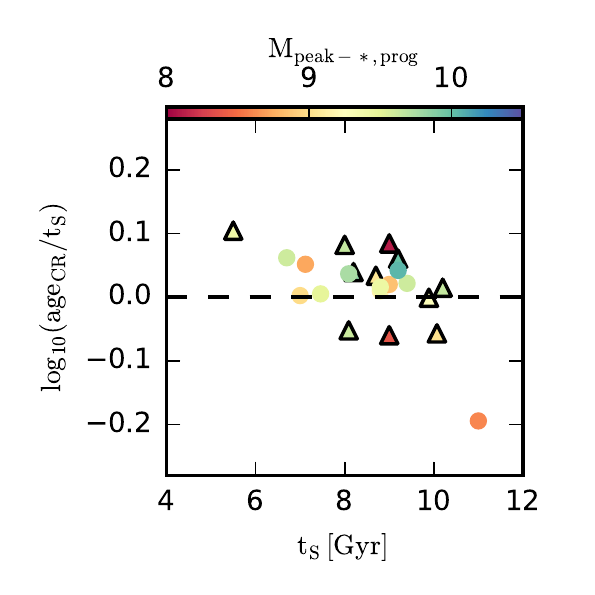}
\caption{{\it Left and middle panels:} the counter-rotating fraction of local stars with $\rm[\alpha/Fe]>0.1$ and $-1.2 < \rm[Fe/H] < -0.5$ correlates with the GES progenitor stellar mass (left panel) and the ratio of the stellar mass of the GES progenitor and the host proto-galaxy (middle panel). {\it Right panel:} the 20th percentile of the age distribution of these stars as a function of the true merger time (as defined by \textlcsc{SUBFIND}). The merger time of the GES progenitor is estimated to within a $\sim 10\%$ accuracy for simulations with GES progenitors on radial orbits and a ``Sausage-like'' metal-rich stellar halo.}
\label{diag}
\end{figure*}

In the previous section, we presented and discussed the detailed chemodynamic properties of the proto-galaxy stars and starburst populations for a representative case. In this section, we present the kinematic properties of these populations for all 22 simulations considered in this work. This set of cosmological simulations samples the natural variation of proto-galaxy properties and merger properties expected in the $\Lambda$CDM cosmological paradigm, and therefore affords us the opportunity to acquire a statistical handle on how the GES merger event shaped the Galaxy. We explore the correlation between dynamical heating of the proto-galaxy with the GES progenitor mass, and discuss potential diagnostics to weigh and date the GES merger, that, while challenging, may help future observations. Furthermore, we explore the dependency of these relations on the orbital eccentricity of the last significant merger in each simulation. For this analysis, we select only local star particles ($11 > R > 5$ kpc, $|Z| < 3$ kpc) which elucidates our results and provides a more practical prediction for corroboration by current Galactic surveys. 

\subsubsection{Dependency on kinematics, GES mass and eccentricity}

In Fig.~\ref{ksum}, we compare the vertical velocity dispersion (left panel), rotational velocity (middle panel) and the ratio of the rotational velocity to the vertical velocity dispersion (right panel) of the starburst population to that of the proto-galaxy population. From this figure, we see that the starburst population is in most haloes kinematically cooler than the proto-galaxy component, in some cases by more than a factor of two (if the $v_{\phi}/\sigma _z$ ratio is considered). We note that the majority of simulations have built a proto-disc ($v_{\phi}/\sigma _z >1$) before the GES merger occurs. However, the proto-galaxy and starburst populations at present day have a range of kinematic properties; many simulations have $ v_{\phi}/\sigma _z <1$, indicating a degree of dynamical heating. Interestingly, the kinematics of these populations appears to correlate with the mass of the GES progenitor: low values of $v_{\phi}/\sigma _z$ appear to have had more massive mergers.

Motivated by the variation of the kinematics of the thick disc populations in our simulations, we explore links between the properties of the GES merger and local proto-galaxy stars (stars that were born in-situ in a cylindrical annulus around the solar radius before the GES merger time). We first calculate the radial and vertical velocity dispersion of local proto-galaxy stars before the merger event and at present-day, and show the fractional change of these quantities as a function of their initial velocity dispersion and GES progenitor stellar mass, in the top-left and bottom-left panels of Fig.~\ref{heatsum}, respectively. Here, the GES progenitor mass is taken to be the maximum stellar mass that the progenitor reached before merging, which typically translates to its stellar mass before infall. 

The top panels of this figure indicate that the fractional increase in velocity dispersion (in both the vertical and radial directions) is anti-correlated with the velocity dispersion of the proto-galaxy prior to the GES merger; kinematically colder proto-galaxy/disc configurations experience a larger fractional amount of kinematic heating\footnote{In Fig.~\ref{heatsumabs}, we show that the absolute kinematic heating in both directions shows no clear correlation with the pre-merger velocity dispersion of the proto-galaxy.}. The middle column of Fig.~\ref{heatsum} shows that the fractional kinematic heating is proportional to the stellar mass of the GES progenitor also. The scatter in this relation is (in part) driven by the kinematic state of the proto-galaxy prior to the merger: for fixed GES progenitor mass (e.g. $\sim 10^9$ $\rm M_{\odot}$), haloes in which the proto-disc has a lower velocity dispersion experience a greater amount of \emph{fractional} kinematic heating (although this is not the case for the \emph{absolute} heating; see Fig.~\ref{heatsumabs}) compared to those of a higher velocity dispersion, reflecting the trend in the left panels of the figure. 

In addition to the merger mass, it is perhaps natural to expect that the transfer of energy to the proto-galaxy from the merger depends also on the orbital eccentricity of the merger; in-spiraling mergers transfer energy more slowly compared to highly radial mergers with low impact parameters. Moreover, mergers that gradually sink into the centre of the proto-galaxy tend to do so in co-planar orbits owing to disc-torquing \citep{GGM17b}, and therefore transfer energy differently compared to mergers on radial orbits. To determine how this affects the kinematic heating imparted to the proto-galaxy, we calculate the orbital eccentricity of the GES progenitor prior to merging as a function of fractional change in vertical and radial velocity dispersion (right panels in Fig.~\ref{heatsum}). The variation in the pre-merger proto-disc kinematics introduces scatter to this relation, as for the correlation found for the progenitor mass in the middle panels. Excluding extreme values for the pre-merger proto-disc velocity dispersion reveals a mild positive correlation, which indicates that mergers on radial orbits tend to provide more kinematic heating to the proto-galaxy than do those on more tangential orbits (this trend is also present in terms of absolute kinematic heating; see Fig.~\ref{heatsumabs}). This can be clearly seen in the middle panels of Fig.~\ref{heatsum} for pre-merger proto-galaxy velocity dispersions of $\sigma _{\rm rb}\sim 80$ $\rm km \,s^{-1}$ and $\sigma _{\rm zb}\sim 60$ $\rm km \,s^{-1}$; the mergers with more radial orbits lie above those with more tangential orbits (in the majority) for a range of progenitor masses. Thus, we conclude that, because of its radial orbit, the GES progenitor likely provided more kinematic heating to the proto-galaxy compared to a more tangentially biased merger of the same mass.

\subsubsection{Constraints on the mass and time of the GES merger}

That the kinematic heating of the proto-galaxy/Splash population is proportional to the mass of the progenitor suggests that an observational indicator of the progenitor mass could be the number (or fraction) of pre-existing stars kicked out into the stellar halo \citep[as discussed in][]{BSF20}. It follows that the fraction of proto-galaxy stars on counter-rotating orbits may correlate with the GES progenitor mass. Motivated by the distribution of the Splash stars in the $\rm[\alpha/Fe]$-$\rm[Fe/H]$ plane for the simulation discussed in the previous section (Fig.~\ref{aferz}), we make a selection of counter-rotating, in-situ stars with $\rm[\alpha/Fe]>0.1$ and $-1.2 < \rm[Fe/H] < -0.5$ as an observational proxy for Splash stars. Note that these values will be different for the Milky Way, and are intended only to provide a rough idea of the accuracy with which the GES merger may be dated with this method. In the left panel of Fig.~\ref{diag}, we show the counter-rotating fraction of these stars as a function of the GES progenitor mass. A clearly positive correlation is produced, with some scatter from the varying pre-merger kinematic state of the proto-galaxies. The proto-galaxy in a small number of simulations have notably lower counter-rotating fraction for their GES progenitor mass compared to the general trend, which is because their proto-galaxy at the GES merger time is relatively massive: the middle panel of Fig.~\ref{diag} shows that they shift to lower GES-proto-galaxy stellar mass ratios compared to the others. It is important to note that mergers that precede the GES-like merger contribute also to the fraction of counter rotating stars, which will introduce scatter to this relation. Therefore, the early merger history of the Milky Way in addition to the mass growth of the proto-galaxy are important pieces of information required to fully decipher this trend. Nevertheless, we predict that the fraction of counter-rotating proto-galaxy stars places constraints on the mass of the GES progenitor. However, we caution that application to the Milky Way data will require the careful separation of in-situ proto-halo/Splash stars from accreted stars.

The dynamical connection between the GES merger and the counter-rotating Splash population offers the possibility to date the GES merger time from the age distribution of these stars. We show the 20th percentile of the age distribution of our selected stars against the true GES merger time for all our simulated haloes in the right panel of Fig.~\ref{diag}. With the exception of one simulation, the GES merger time may be inferred to within $25\%$ accuracy in all cases. The accuracy is even better (within $10\%$) for the GES progenitors on radial orbits, which indicates that the merger time of $\sim 9.5$ Gyr ago derived by \citet{BSF20}, which is consistent with that derived by \citet{DiMatteo+Haywood+Lehnert+18}, is robust to within 1 Gyr. Finally, we note that if stellar ages of old stars can be measured accurately, then the age of the starburst induced by the GES merger should provide further constraints on the GES merger time.

\section{Discussion and Conclusions}
\label{conclusions}

We analyse a set of 22 cosmological MHD simulations for the formation of Milky Way-mass galaxies in the context of the GES progenitor and its effects on the formation of the stellar halo and thick disc. These simulations have varying degrees of velocity anisotropy in their stellar halo populations: 9 have been identified to have a ``Sausage-like'' feature similar to that found in the Milky Way \citep{FBD19}. For each simulated galaxy, we split the stellar population at redshift zero into stars formed in the proto-galaxy before the GES merger event and an additional in-situ population induced by a starburst during the merger of the GES progenitor. Our study focused on how to relate this ``mash'' of stellar populations to the stellar disc and halo component of Milky Way-sized galaxies through their chemodynamical properties, and to understand how we can use this information to make predictions for the mass and the merger time of the GES progenitor system. 

Our findings indicate that the GES merger has a two-fold effect on the formation of the thick disc:

\begin{itemize}
\item{} it dynamically heats the proto-galaxy stars, such that some proto-disc stars are scattered into the halo (the ``Splash'') and the proto-disc as a whole becomes more radially extended; 
\item{} the GES merger is in all our simulations gas-rich, and is always associated with a (centrally concentrated) starburst that peaks strongly around the time of merging. The GES progenitor contributes around 10-50$\%$ of the gas consumed by this starburst.
\end{itemize}

After the starburst, the thin disc formation begins in earnest to form upside-down and inside-out from the gradual accretion and settling of relatively metal-poor gas \citep[as demonstrated by][]{GBG18}. This forms the now-familiar picture of a centrally concentrated, chemically-defined thick-disc embedded in a radially-extended, flaring thin disc \citep{MMS15,GSG16,BRS15,KGG17}. 

Our case study of the chemodynamical properties of these components in an individual galaxy reveal that: 

\begin{itemize}
\item{} the merger-induced starburst produces a rotationally supported and compact component (starburst-disc component) that occupies a region in chemo-kinematic space very similar to that of the traditional thick disc. A spheroidal component is created also during the starburst owing to the hydrodynamical pressure that pushes dense star-forming gas onto non-circular, halo-like orbits. 
\item{} the proto-disc component extends across a wide range of chemical abundance space; from the metal-poor, high-alpha region down to the relatively metal-rich starburst thick disc. The locus of this population in abundance space appears to be roughly constant at different locations in real space. The halo component of the proto-galaxy stars (including the Splash) has similar chemistry to proto-disc stars in the inner regions, but is more metal-poor in the solar neighbourhood and beyond. 
\item{} The rotation velocity-metallicity relation shows a positive correlation in both the proto-disc and proto-halo components. We show that this is because at very early epochs prior to the GES merger, the proto-disc grows more massive, rotationally supported and more metal-rich as minor mergers bombard it, creating a relatively metal-poor in-situ Splash population. The GES merger scatters more metal-rich, rotationally supported stars into the halo, and naturally develops a positive relationship between rotation velocity and metallicity in both the proto-disc and proto-halo. The creation of the relatively metal-rich and highly rotating starburst-thick disc component reinforces this trend in our dual-origin formation scenario of the thick disc; an alternative scenario to the typically proposed inside-out formation of the thick disc.
\end{itemize}

Our analysis of the proto-galaxy and starburst populations, and the GES progenitor that shaped them, reveals the following:

\begin{itemize}
\item{} The proto-disc population in the majority of simulations has a $v/\sigma >1$ before the GES progenitor merges, and are heated by up to a factor of 2 in the radial and vertical directions, reducing their fraction of rotational energy. The proto-galaxy is very hot in a minority of simulations.
\item{} The degree of \emph{fractional} kinematic heating caused by the GES progenitor correlates with: i) progenitor mass (more massive progenitors provide more kinematic heating); ii) orbital eccentricity of the GES progenitor (more eccentric orbits lead to more kinematic heating); and iii) the pre-merger proto-galaxy kinematics (colder proto-discs preferentially experience more fractional kinematic heating relative to hotter proto-galaxy configurations). The amount of absolute kinematic heating is primarily a function of i) and ii), with no clear dependence on the initial kinematics of the proto-galaxy(-disc).
\item{} The number/fraction of counter rotating Splash stars in the local vicinity seems to positively correlate with the stellar mass and stellar mass ratio of the GES progenitor. The age distribution of this population appears to be predict the GES merger time to within about $10\%$, and is thus a promising observational diagnostic.
\end{itemize}

The chemodynamic trends of the proto-disc and proto-halo and of the starburst disc are consistent with recent observations that interpret the thick disc/halo to be smoothly connected \citep{DiMatteo+Haywood+Lehnert+18}. Our results therefore demonstrate that the observed properties of these components can be borne out by the hierarchical merging of gas-rich subhaloes that both create and perturb generations of in-situ stars of increasing metallicity as time proceeds. The dominance of the relatively metal-rich stars of the Splash may be explained as the older, metal-poorer populations receiving more random kinetic energy from repeated mergers, which moves them mainly onto extended, halo-like orbits (Fig.~\ref{aferz} and Fig.~\ref{metvrot}). As the disc becomes heavier and more enriched, the relatively large-dynamical impact of the GES merger scatters more metal-rich stars into the halo. The net result is a prominent metal-rich halo and a paucity of stars identified as metal-weak thick disc stars \citep{CCI20} in the local vicinity, which is subdominant to the traditional thick disc. The dominance of metal-rich thick disc over the relatively metal-poor thick disc/halo is further enhanced by the starburst material, which contributes a similar mass to the metal-rich thick disc as proto-disc stars (Table~\ref{t1}). Besides the thick disc, the starburst induced by the GES merger generates an in-situ spheroidal component of stars. This component may represent a formation pathway for globular clusters, because it provides the high-pressure environment and rapid star-formation timescales necessary to create clusters with high-alpha abundances in the central regions of the galaxy \citep{Minniti95,Pritzl+Venn+Irwin2005}.

We have shown that the properties of the Splash component are linked to those of the GES progenitor. Although several factors such as the proto-galaxy chemodynamic properties, GES progenitor orbit, and the merger history prior to the GES merger all play a role in shaping the Splash population, we showed that the counter-rotating fraction of proto-galaxy stars can in principal be used to infer the stellar mass (ratio) of the GES progenitor. Using the EAGLE simulations to interpret the observed stellar halo chemodynamics, \citet{Mackereth+Schiavon+Pfeffer+19} estimated the GES progenitor stellar mass to be $< 10^9$ $\rm M_{\odot}$, lower than the recent total stellar halo mass estimate of $\sim1.4\times 10^9$ $\rm M_{\odot}$ \citep{DBS20}, and consistent with the idea that the stellar halo is dominated by the GES progenitor. Interestingly, this GES progenitor mass corresponds to less than $5\%$ of the Milky Way mass (at the time the GES progenitor reached its maximum stellar mass) in our sample of simulations, which is much less than the originally claimed mass ratio of around 1:4 \citep{HBK18}. We note that such a small stellar mass ratio has been shown to be conducive to the formation of a boxy-peanut bulge with chemo-dynamical properties compatible with the Milky Way, in contrast to larger mass ratios that seem to provide too much dynamical heating for the required dynamical instability to develop in our set of galaxies \citep[see][for a compelling investigation]{Fragkoudi+Grand+Pakmor+19}. It is also interesting to note that \citet{MGG19} found that stellar haloes analogous to that of the Milky Way should form from a number of low-mass progenitors in addition to the GES progenitor. In our case study, these lower-mass mergers occurred prior to the GES merger event, which we have explained to be important for shaping the chemodynamical trends of the Splash population. 

Finally, we remark that the ages of counter rotating stars in the local vicinity make a reliable estimate for the time of the GES merger event, which has been found in \citet{BSF20} to be about 9.5 Gyr ago. Given the clear connection between the GES merger and a starburst of short duration, the identification of a clear peak in the SFH at early times may provide a good estimate of the merger time as well \citep[see the recent works of][that find evidence for an early burst of star formation in the disc]{NSG20,FNP20}. As stellar age measurements improve in the future \citep[e.g.][]{CKM20,MCM20}, we can expect to narrow down the time at which the last significant merger in the Milky Way occurred. 

\section*{Acknowledgements}
We thank the anonymous referee for providing a helpful and constructive report. This project was developed in part at the 2019 Santa Barbara ``Dynamical models for stars and gas in the Gaia Era'' program, hosted by the Kavli Institute for Theoretical Physics at the University of California, Santa Barbara. FAG acknowledges financial support from CONICYT through the project FONDECYT Regular Nr. 1181264, and funding from the Max Planck Society through a Partner Group grant. FM acknowledges support through the program ``Rita Levi Montalcini'' of the Italian MIUR.

\section*{Data Availability}
The data underlying this article will be shared on reasonable request to the corresponding author.

\bibliographystyle{mnras}
\bibliography{mnras_template.bbl}

\appendix
\section{Absolute kinematic heating of the proto-galaxy}
\label{app}

\begin{figure*}
\includegraphics[scale=1.25,trim={0.cm 0.1cm 0.4cm 0.2cm},clip]{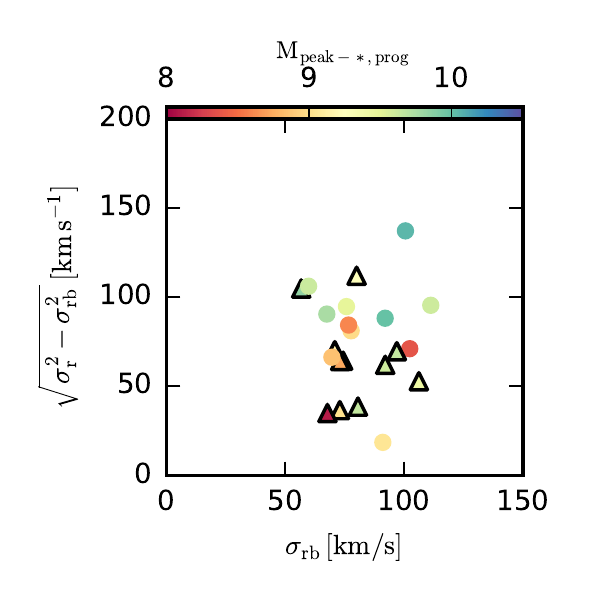}
\includegraphics[scale=1.25,trim={1.6cm 0.1cm 0.4cm 0.2cm},clip]{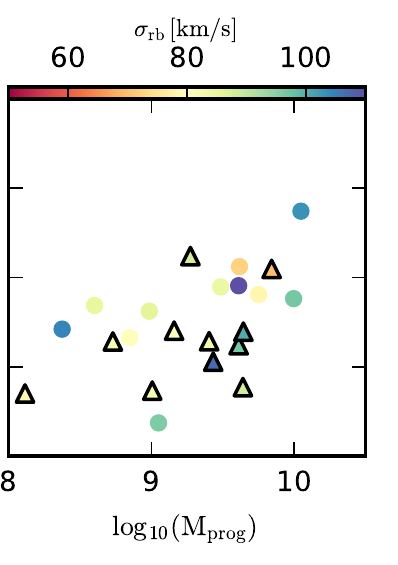}
\includegraphics[scale=1.25,trim={1.6cm 0.1cm 0.4cm 0.2cm},clip]{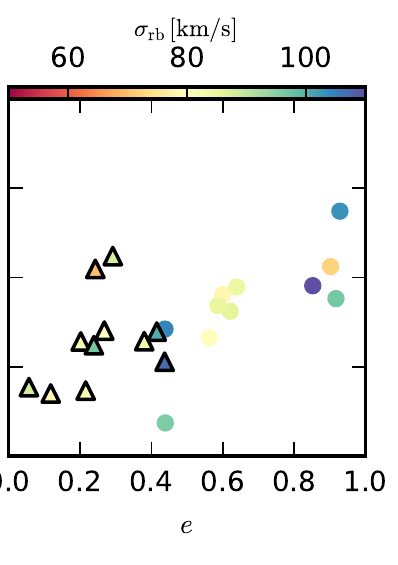}\\
\includegraphics[scale=1.25,trim={0.cm 0 0.4cm 0.2cm},clip]{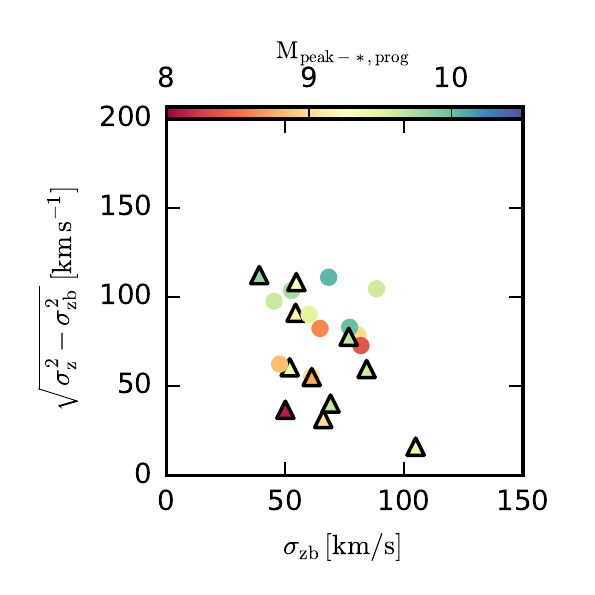}
\includegraphics[scale=1.25,trim={1.6cm 0 0.4cm 0.2cm},clip]{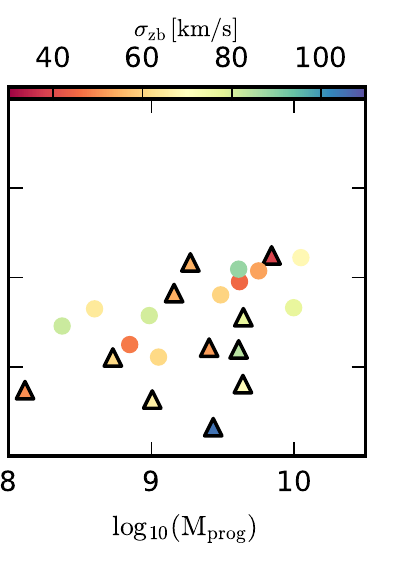}
\includegraphics[scale=1.25,trim={1.6cm 0 0.4cm 0.2cm},clip]{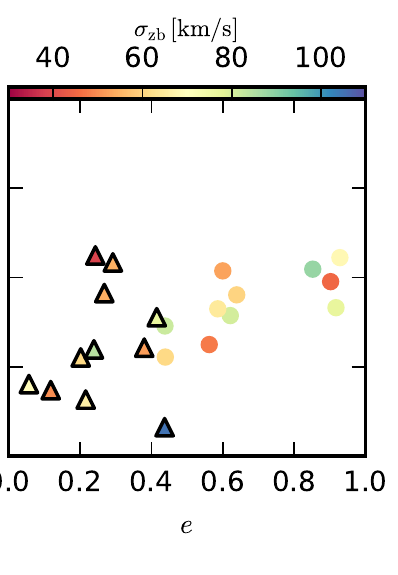}
\caption{As Fig.~\ref{heatsum}, but for the absolute changes (unnormalised) in radial and vertical velocity dispersion.}
\label{heatsumabs}
\end{figure*}

Fig.~\ref{heatsumabs} is analogous to Fig.~\ref{heatsum}, but explores correlations between the absolute kinematic heating imparted to the proto-galaxy by the GES merger, in contrast to the relative kinematic heating. This figure shows that the dependency of absolute heating on the initial velocity dispersion of the proto-galaxy is weaker compared to the dependency on the stellar mass and orbital eccentricity of the GES progenitor. 

\bsp	
\label{lastpage}
\end{document}